\documentclass{jfm}
\usepackage{lineno}
\usepackage{color}
\usepackage{graphicx}
\usepackage{booktabs}
\usepackage[x11names, svgnames, dvipsnames]{xcolor}
\definecolor{Burgundy}{RGB}{144,0,32}
\usepackage{subfig}
\usepackage{soul} 
\usepackage{layouts}
\usepackage{lscape}
\usepackage{mathtools}
\usepackage{relsize} 
\usepackage{xcolor}
\usepackage{inputenc}
\usepackage{floatrow}
\usepackage{natbib}
\usepackage{amsmath}
\usepackage{commath}
\usepackage{enumitem}
\graphicspath{{./figures/}}

\showboxdepth=\maxdimen
\showboxbreadth=\maxdimen

\newcommand*\diff{\mathop{}\!\mathrm{d}}
\newcommand*\diag{\mathop{}\!\mathrm{diag}}
\newcommand*\tke{\mathop{}\!\mathrm{TKE}}	

\newcommand\Fro{\mbox{\textit{Fr}}}  
\newcommand\Str{\mbox{\textit{St}}}  
\newcommand{\appropto}{\mathrel{\vcenter{
\offinterlineskip\halign{\hfil$##$\cr
\propto\cr\noalign{\kern2pt}\sim\cr\noalign{\kern-2pt}}}}}
\newcommand{\GG}[1]{}

\shorttitle{SPOD analysis of stratified wakes}
\shortauthor{Nidhan et al.}

\title{
Analysis of coherence in turbulent stratified wakes using spectral proper orthogonal decomposition}

\author{Sheel Nidhan\aff{1},
Oliver T. Schmidt\aff{1}
\and Sutanu Sarkar\aff{1}  \corresp{\email{sarkar@ucsd.edu}}}

\affiliation{\aff{1}Department of Mechanical and Aerospace Engineering, University of California San Diego, CA 92093, USA}

\begin{document} 

\maketitle
\floatsetup[figure]{subcapbesideposition=top}
\floatsetup[subfigure]{subcapbesideposition=top}

\begin{abstract}		

We use spectral proper orthogonal decomposition (SPOD) to extract and analyze coherent structures in the turbulent wake of a disk at  
Reynolds number $\Rey = 5 \times 10^{4}$ and Froude numbers $\Fro$ = $2, 10$. We find that the SPOD eigenspectra of both wakes exhibit a low-rank behavior and  the relative contribution of low-rank modes to total fluctuation energy increases with $x/D$. The vortex shedding (VS) mechanism, which corresponds to $\Str \approx 0.11-0.13$ in both wakes, is active and dominant throughout the domain in both wakes. The continual downstream  decay of the SPOD eigenspectrum peak at the VS mode, which is a prominent feature of the unstratified wake, is inhibited by buoyancy, particularly for $\Fro = 2$. The energy at and near the VS  frequency is found to appear in the outer region of the wake when the downstream distance exceeds $Nt = Nx/U =  6 - 8$. Visualizations show that  unsteady internal gravity waves (IGWs) emerge at the same $Nt =  6 - 8$. A causal link between the VS mechanism and the unsteady IGW generation is also established using the SPOD-based reconstruction and analysis of the pressure-transport term. These IGWs are also picked up in SPOD analysis as a structural change in the shape of the leading SPOD eigenmode. The $\Fro = 2$ wake shows layering in  the wake core at {$Nt > 15$} which is captured by the leading SPOD eigenmodes of the VS frequency at downstream locations $x/D > 30$. The VS mode of the $\Fro = 2$ wake is streamwise-coherent, consisting of V-shaped structures at $x/D \gtrsim 30$. Overall, we find that the coherence of wakes, initiated by the VS mode at the body, is  prolonged by buoyancy to far downstream.  Also, this coherence is spatially modified by buoyancy into horizontal layers and IGWs. Low-order truncations of SPOD modes are  shown to efficiently reconstruct important second-order statistics.

\end{abstract}

\begin{keywords}
Authors should not enter keywords on the manuscript, as these must be chosen by the author during the online submission process and will then be added during the  typesetting process (see http://journals.cambridge.org/data/\linebreak[3]relatedlink/jfm-\linebreak[3]keywords.pdf for the full list)
\end{keywords}

\section{Introduction}\label{section_introduction}
Turbulent wakes are ubiquitous both in nature and man-made devices. From flow past moving vehicles \citep{grandemange_study_2015} to flow past topographic features \citep{puthan2021tidal} in oceans, they play an important role in transporting momentum and energy across large distances from the wake generator.  In 
the ocean and the atmosphere, the background density often has a stable density stratification.  
Buoyancy in a stable background enables  the emergence of several distinctive features, e.g.,
suppression of vertical turbulent motions \citep{spedding_vertical_2002}, multistage wake decay \citep{lin1979wakes,spedding_evolution_1997}, appearance of coherent structures in the late wake \citep{lin1979wakes,lin_turbulent_1992}, and formation of steady~\citep{hunt_experiments_1980} and unsteady \citep{gilreath_experiments_1985,bonneton_internal_1993} internal gravity waves, to name a few. A majority of wake studies utilize axisymmetric body shapes (sphere, disk,  spheroid, etc.) since such canonical shapes make it convenient to understand the phenomenology of turbulent stratified wakes. 

The existence of coherent structures has been established to be an universal feature of both unstratified and stratified turbulent wakes. The K\'arman vortex street associated with vortex shedding from the body at a specific frequency is a well known feature of unstratified bluff body wakes which arises from the global instability of  the $m =1$ azimuthal mode as was demonstrated for a sphere by \cite{natarajan_instability_1993} and \cite{tomboulides2000numerical}. The Strouhal number ($\Str$) associated with vortex shedding varies  with the shape of the body. 
Vortex shedding has been investigated  in stratified wakes too. \cite{lin_stratified_1992} conducted a detailed experimental investigation of stratified flow past a sphere of diameter $D$ towed with speed $U$ in a fluid with buoyancy frequency $N$ for $5 \leq \Rey \ (UD/\nu) \leq 10^{4}$ and $0.005 \leq  \Fro \ (U/ND) \leq 20$. At $\Fro \gtrsim 2$, they found that $\Str$  in the near wake of the sphere, at $x/D \approx 3$, attained a constant value of $\Str \approx 0.18$, same as in the unstratified wake. For $\Fro \lesssim 2$, the vortex shedding was two-dimensional and $\Str$ increased with decreasing $\Fro$ in the near wake, similar to the trend in the flow past a circular cylinder. \cite{chomaz_structure_1993} identified four regimes, differentiated by the value of $\Fro$,  in the near wake of a sphere. These regimes  showed structural differences in the shed vortices and their interactions with the lee wave field. 

Another distinctive feature of the stratified wakes is the generation of IGWs which are of two types: (i) body generated steady lee waves and (ii) wake generated unsteady IGWs.  In their pioneering work on the wake of a self-propelled slender body, \cite{gilreath_experiments_1985} noted a  coupling between the unsteady IGWs in the outer wake and the wake core turbulence, which suggests that the generation of the unsteady IGWs is inherently nonlinear in nature. \cite{bonneton_internal_1993} and \cite{bonneton1996structure} examined IGWs in the flow past a sphere. Lee waves were found to dominate when $\Fro \lesssim 0.75$ and, for $\Fro \gtrsim 2.25$, the downstream wake was dominated by the unsteady IGWs. Analysis of the density and velocity spectra in the outer wake showed a distinct peak at the vortex shedding frequency of sphere, $\Str \approx 0.18$. \cite{brandt_internal_2015} found wake turbulence to be a dominant source term for IGWs at $\Fro \gtrsim 1$ in their experimental work on sphere wakes. However, they did not expand on the spectral characteristics of these wake-generated IGWs. Recently, \cite{meunier2018internal} also conducted a theoretical and experimental study of waves generated by various wake generators, focusing primarily on the scalings of wavelengths and amplitudes across various $\Fro$ and wake generators. Various aspects of IGWs have also been studied through numerical simulations \citep{abdilghanie_internal_2013,zhou2016surface,ortiz-tarin_stratified_2019,rowe_internal_2020}.

In the last two decades, the rise in computing power has enabled a number of numerical studies which have improved our understanding of stratified wakes. A large body of numerical literature employs the temporal model wherein the wake generator is not included \citep{gourlay_numerical_2001,dommermuth_numerical_2002,brucker_comparative_2010,diamessis_similarity_2011,de_stadler_simulation_2012,abdilghanie_internal_2013,redford_numerical_2015,zhou_large-scale_2019,rowe_internal_2020}. Instead, these simulations are initialized with synthetic mean and turbulence profiles mimicking those of a wake. Body-inclusive simulations which resolve the flow at the wake generator and at a high enough $\Rey$ that sustain turbulence are relatively recent \citep{orr_numerical_2015,pal_regeneration_2016,pal_direct_2017,ortiz-tarin_stratified_2019,chongsiripinyo_decay_2020}.

The database from the body-inclusive simulation of \cite{chongsiripinyo_decay_2020}, hereafter referred to as CS2020, will be interrogated in this paper to analyze spatio-temporal coherence. CS2020 perform  large eddy simulation (LES)  of flow past a disk   at $\Rey = 5 \times 10^4$ and at various values of $\Fro$. The authors find that the wake transitions through three different regimes of stratified turbulence (provided buoyancy Reynolds number $> O(1)$), each with distinctive  turbulence properties:  weakly stratified turbulence (WST) which commences  when the turbulent Froude number $Fr_h$ decreases to $O(1)$,  intermediately stratified turbulence (IST) when $Fr_h$ decreases to $O(0.1)$, and strongly stratified turbulence (SST) when $Fr_h$ reduces to to $O(0.01)$. Here $\Fro_h = u_h'/Nl_{v}$, where $u'_h$, $N$, and $l_v$ are r.m.s. horizontal velocity fluctuations, buoyancy frequency, and a characteristic turbulent vertical lengthscale, respectively. In the WST regime, the turbulence is not yet appreciably affected by buoyancy effects. Anisotropy in turbulent velocity components, which is a key manifestation of stratification, has not kicked in yet (see figure 8 of CS2020). As the flow evolves downstream, turbulence anisotropy keeps increasing and the turbulence transitions to the IST regime at $\Fro_h \sim O(0.1)$. The SST regime, which commences at $\Fro_h \approx 0.03$, is characterized by a strong anisotropy in turbulence. An indication of arrival of this regime is the scaling of the vertical lengthscale with $u'_h/N$ as derived by \cite{billant_self-similarity_2001} (also see figure 12 in CS2020). In the SST regime, mean defect velocity and $u'_h$ decay at the same rate of $x^{-0.18}$ while vertical turbulent velocity ($u'_z$) decays at a faster rate of $x^{-1}$. Regime classification based on turbulence instead of mean velocity was introduced in the context of stratified homogeneous turbulence, e.g., \cite{brethouwer_scaling_2007}, and was recently extended to stratified turbulent wakes by \cite{zhou_large-scale_2019} and CS2020.

With the huge amount of numerical and experimental data becoming available, data-driven modal decomposition techniques have also seen an unprecedented rise in their use to understand the dynamics and role of coherent structures in turbulent flows. These techniques have also been used to construct reduced-order models of these flows. 
One popular  technique is  proper orthogonal decomposition (POD), proposed by \cite{lumley_1967,lumley_1970} in the context of turbulent flows, which 
provides a set of modes ordered hierarchically in terms of energy content. Another popular technique is dynamic mode decomposition (DMD),  described by \cite{schmid_dynamic_2010}, which decomposes the flow into a set of spatial modes, each oscillating at a specific frequency. 

However, applications of modal decomposition to stratified flows are few in number.
\cite{diamessis_spatial_2010} performed snapshot POD
(\cite{sirovich1987turbulence}) 
on the vorticity field from a temporal simulation
at $\Rey = 5 \times 10^3$ and $\Fro = 2$, noting a link between wake core structures and the angle of emission of IGWs in the outer wake. The layered wake core structure, which is a distinctive feature of stratified turbulent wakes, was found in the POD modes with lower modal index (corresponding to higher energy). As the modal index increased, the wake core was found to be dominated by small-scale incoherent turbulence. \cite{xiang_dynamic_2017} performed spatial and temporal DMD on  the experimental data of the stratified wake of a grid showing that DMD modes successfully captured  lee waves and Kelvin-Helmholtz (KH) instability in the near wake ($Nt < 10$). \cite{nidhan_dynamic_2019} performed three-dimensional  (3D) and planar two-dimensional (2D) DMD on the sphere wake at $\Rey = 500$ and $10^4$, respectively. At $\Rey = 500$ and  $\Fro = 0.125$, they found that the 2D vortex shedding in the center-horizontal plane and `surfboard' structures in the center-vertical plane corresponded to the same DMD mode oscillating at the vortex shedding frequency of $\Str \approx 0.19$. At the higher $\Rey = 10^4$, DMD modes associated with vortex shedding  showed IGWs in the outer wake.

In the present work, we use spectral proper orthogonal decomposition (SPOD), originally proposed by \cite{lumley_1967,lumley_1970} and recently revisited by \cite{towne_spectral_2018}, to identify and analyze the coherent structures in the turbulent stratified wake of a disk at $\Rey = 5\times10^4$. 
In its original form, POD is prohibitively expensive to apply on  today's large numerical databases with high space-time resolution. The form put forward by \cite{towne_spectral_2018} leverages the temporal symmetry of statistically stationary flows to improve computational tractability.
SPOD decomposes statistically stationary flows into energy-ranked modes with monochromatic frequency content, thus separating both the temporal and spatial scales in the flow, unlike the popular snapshot variant given by \cite{sirovich1987turbulence}. SPOD has been used extensively in recent times for analysis of coherent structures and reduced-order modeling in a variety of unstratified flow configurations: (i) turbulent jets \citep{semeraro_stochastic_2016,schmidt_wavepackets_2017,schmidt_spectral_2018,nogueira_large-scale_2019,nekkanti2021modal}, (ii) turbulent wakes \citep{nidhan_spectral_2020}, (iii) channel \citep{muralidhar_spatio-temporal_2019} and pipe \citep{abreu2020spectral} flows, (iv) flow reconstruction \citep{nekkanti2020frequency} and low-order modeling \citep{chu_stochastic_2021}, (v) wakes of actuator disks in turbulent environments \citep{ghate_interaction_2018,ghate_broadband_2020}, etc.

The formation of  coherent  pancake vortices in the Q2D late wake  does not necessarily require vortex shedding from the body as was demonstrated by \cite{gourlay_numerical_2001} whose temporally evolving  model at $\Fro = 10$ did not include the vortex shedding mode but  still exhibited Q2D-regime pancake vortices. Our interest is also in coherent structures but in a region of the far wake which is at large $x/D$ but still not in the Q2D regime. We ask how does buoyancy affect the space-time coherence as the flow  progresses from the near wake to the far wake? What are the salient differences between the unstratified ($\Fro = \infty$) and stratified wakes in the context of coherent structures? We will address these questions by analyzing the LES dataset of CS2020, specifically the wakes at  $\Fro = 2$ and 10. We adopt SPOD for the data analysis since it is well suited to extract modes which have spatial and temporal coherence and thus track the evolution of specific modes, e.g. the vortex shedding (VS) mode, as the wake evolves downstream. The SPOD analysis also allows us to address a second set of questions: (i) are coherent modes linked to unsteady IGWs and (ii) how is the energy in dominant coherent structures distributed across the wake cross-section during downstream evolution? SPOD modes can also be useful for constructing reduced-order models prompting the third question: what is the efficacy of different SPOD modal truncations in regard to the reconstruction of various second-order turbulence statistics in turbulent stratified wakes? 

The rest of the paper is organized as follows.  \secref{section_numerical_methodology} and \secref{section_spod} give a brief overview of the numerical methodology and SPOD technique. Visualizations of $\Fro = 2$ and $10$ wakes are presented in \secref{section_flow_viz}. The characteristics of SPOD eigenvalues and eigenspectrum are discussed in \secref{section_results_eigenvalues_eigenspectra}. The VS mode and its link to the unsteady IGWs are discussed in detail in \secref{section_VSmode}. Sections \secref{section_slice_spod_eigenmodes} and \secref{section_slice_spod_reconstruction} discuss the spatial structure of SPOD eigenmodes and  trends in the reconstruction of  second-order statistics by sets of truncated SPOD modes, respectively. Finally, the discussion and conclusions are presented in \secref{section_conclusions}.

\section{Numerical methodology}\label{section_numerical_methodology}
We use the numerical database of the wake of a circular disk at $\Rey$ = $5 \times 10^4$ from CS2020.
In particular, we analyze the datasets of stratified wakes at $\Fro = 2$ and $10$ from their numerical database. CS2020 use high-resolution large eddy simulation (LES) to numerically solve the filtered Navier-Stokes equations  system along with density diffusion equation under the Boussinesq approximation. 

These equations are as follows:

continuity,
\begin{equation} 
	\frac{\partial u_{i}}{\partial x_{i}} = 0,
	\label{conservation_eqn}
\end{equation}

momentum,

\begin{equation} 
	\frac{\partial u_{i}}{\partial t} + \frac{\partial (u_{i}u_{j})}{\partial x_{j}} = -\frac{\partial p}{\partial x_{i}} + \frac{1}{\Rey}\frac{\partial}{\partial x_{j}}\Big[\Big(1 + \frac{\nu_{s}}{\nu}\Big)\frac{\partial u_{i}}{\partial x_{j}}\Big] - \frac{1}{\Fro^{2}}\rho'\delta_{i3}, 
	\label{momentum_eqn}
\end{equation}

and density diffusion,

\begin{equation}
	\frac{\partial \rho}{\partial t} + \frac{\partial (\rho u_{j})}{\partial x_{j}} = \frac{1}{\Rey \Pran}\frac{\partial}{\partial x_{j}}\Big[\Big(1 + \frac{\kappa_{s}}{\kappa}\Big)\frac{\partial \rho}{\partial x_{j}}\Big],
	\label{density_eqn}
\end{equation}

where $u_{i}$ corresponding to $i=1, 2,$ and $3$ refer to velocity in the streamwise ($x_{1}$ or $x$), lateral ($x_{2}$ or $y$), and vertical ($x_{3}$ or $z$) directions, respectively. Gravity acts in the vertical direction (\ref{momentum_eqn}). The density field is decomposed into a background profile, $\rho_b(z) = \rho_o + (\diff \rho_b/\diff z)z$ (where $K$ is a constant), and density deviation ($\rho'$). Thus $\rho(x,y,z,t) = \rho_{b}(z) + \rho'(x,y,z,t)$. In (\ref{momentum_eqn}), $\nu_{s}$ and $\nu$ refer to the subgrid kinematic viscosity obtained from LES and kinematic viscosity of the fluid, respectively. Likewise, $\kappa_{s}$ and $\kappa$ in equation (\ref{density_eqn}) refer to the subgrid density diffusivity and  density diffusivity of the fluid, respectively.

(\ref{conservation_eqn}) - (\ref{density_eqn}) are non-dimensionalized using the following parameters: (i) free stream velocity ($U_{\infty}$) for velocity field, (ii) diameter of disk ($D$) for spatial locations $x_{i}$, (iii) dynamic pressure ($\rho_{o}U_{\infty}^{2}$) for pressure field, (iv) advection timescale ($D/U_{\infty}$) for time $t$, and (iv) $-(\diff \rho_b/\diff z)D$ for density deviation. There are three non-dimensional parameters of interest: (1) body-based Reynolds number ($\Rey$) defined as $U_{\infty}D/\nu$, (2) body-based Froude number ($\Fro$) defined as $U_{\infty}/ND$ where $N$ is the buoyancy frequency, $N^{2} = -g/\rho_{o}(\diff \rho_b/\diff z)$, and (3) Prandtl number ($\Pran$) defined as $\nu/\kappa$ which is set as 1 in  CS2020 simulations. $\kappa_{s}$ is also set equal to $\nu_{s}$ for the LES simulations.

A cylindrical coordinate system is adopted and the disk is represented using the immersed boundary method (IBM) of \cite{balaras_modeling_2004,yang_embedded-boundary_2006}. Spatial derivatives are computed using second-order central finite differences  and temporal marching is performed using a fractional step method which combines a low-storage Runge-Kutta scheme (RKW3) with the second-order Crank-Nicolson scheme. The kinematic subgrid viscosity ($\nu_{s}$) and density diffusivity ($\kappa_{s}$) are obtained using the dynamic eddy viscosity model of \cite{germano_dynamic_1991}. At the inlet and outlet, Dirichlet inflow and Orlanski-type convective (\cite{orlanski_simple_1976}) boundary conditions are specified, respectively. The Neumann boundary condition is used at the radial boundary for the density and velocity fields. To prevent the spurious propagation of  internal waves upon reflection from the boundaries, sponge regions with Rayleigh-damping are employed at radial, inlet, and outlet boundaries. 
The radial and streamwise domains span  $0 \leq r/D \leq 80$ and $-30 \leq x/D \leq 125$, respectively. A large radial extent facilitates weakening of the IGWs before they hit the boundary and thereby also controls the amplitude of spurious reflected waves. The distribution of grid points are as follows: $N_{r} = 531$ in the  radial direction, $N_{\theta} = 256$ in the azimuthal direction, and $N_{x} = 4608$ in the streamwise direction, resulting in approximately 530 million elements. The grid resolution is excellent by LES standards in all three directions. Readers may refer to \cite{chongsiripinyo_decay_2020} for more details on the grid resolution and numerical scheme.

\section{Spectral proper orthogonal decomposition - theory and present application}\label{section_spod}
In this work, we employ spectral POD (SPOD) to study the dynamics of coherent structures in stratified wakes, rather than the more commonly employed snapshot POD \citep{sirovich1987turbulence}. SPOD enables the identification of dominant structures evolving coherently in both space and time by exploiting temporal correlation among flow snapshots. This approach is particularly well-suited for flow configurations like turbulent wakes which are known to be dominated by mechanisms operating at specific frequencies, e.g., vortex shedding, pumping of recirculation bubble, shear layer breakdown, to name a few \citep{berger_coherent_1990}. On the contrary, snapshot POD assumes each snapshot of the flow to be an independent realization. As a result, the temporal coherence of POD modes is not guaranteed. Furthermore, it can be also shown that the coefficients dictating the temporal evolution of snapshot POD modes are broadband, i.e., containing contributions from a range of frequencies \citep{towne_spectral_2018}. SPOD requires a larger amount of time-resolved data compared to snapshot POD. Hence, snapshot POD has dominated the literature compared to SPOD.

\subsection{Theory of SPOD for statistically-stationary stratified flows}\label{subsection_theory_spod}

For the SPOD analysis of stratified wakes, the  fluctuating density fields $(\rho'(\mathbf{x},t))$ and velocity fields $(\mathbf{u}'(\mathbf{x},t) = [{u}_{r}'(\mathbf{x},t),{u}_{\theta}'(\mathbf{x},t),{u}_{x}'(\mathbf{x},t)]^{T})$ are taken  together as a single state-space field $\boldsymbol{\Lambda}(\textbf{x},t) = [\mathbf{u}'(\mathbf{x},t), \rho'(\mathbf{x},t)]^{T}$. Following \cite{lumley_1970}, we seek POD modes $\boldsymbol{\Psi}(\mathbf{x},t)$ that have maximum ensembled-average projection on $\boldsymbol{\Lambda}(\mathbf{x},t)$, expressed as: 
\begin{equation}	
	\max_{\boldsymbol{\Psi}}  \frac{\langle |\{\boldsymbol{\Lambda}(\mathbf{x},t), \boldsymbol{\Psi}(\mathbf{x},t)\}|^{2}\rangle}{\{\boldsymbol{\Psi}(\mathbf{x},t),\boldsymbol{\Psi}(\mathbf{x},t)\}},
	\label{pod_maximize}
\end{equation}
where $\langle . \rangle $ denotes the ensemble average. We define the inner product $\{\boldsymbol{\Lambda}^{(1)}(\mathbf{x},t), \boldsymbol{\Lambda}^{(2)}(\mathbf{x},t)\}$ as: 
\begin{equation}
	\{\boldsymbol{\Lambda^{(1)}}(\mathbf{x},t), \boldsymbol{\Lambda^{(2)}}(\mathbf{x},t)\} = \int_{-\infty}^{\infty}\int_{\Omega} \boldsymbol{\Lambda^{(2)}}^{*}(\mathbf{x},t)\diag\Big(1,1,1,\frac{g^{2}}{\rho^{2}_{o}N^{2}} \Big)\boldsymbol{\Lambda^{(1)}}(\mathbf{x},t)\diff\mathbf{x}\diff t,
	\label{inner_product}
\end{equation}
where $(.)^{*}$ denotes the Hermitian transpose. 
The so-defined  inner-product norm ensures that the obtained POD modes are optimal in terms of capturing two-times the overall sum of turbulent kinetic energy (TKE) and turbulent potential energy (TPE),
where TKE $ = \langle u'_{i}u'_{i} \rangle/2$ and TPE=$\frac{g^{2}}{2\rho^{2}_{o}N^{2}}\langle \rho'\rho'\rangle$. 

Following \cite{holmes2012turbulence}, (\ref{pod_maximize}) can be expressed as a Fredholm-type integral eigenvalue problem as follows:
\begin{equation} 
	\int_{-\infty}^{\infty}\int_{\Omega}R_{ij}(\mathbf{x},\mathbf{x}',t,t')\mathbf{W}(\mathbf{x}')\Psi^{(n)}_{j}(\mathbf{x}',t')\diff\mathbf{x}'\diff t' = \lambda^{(n)} \Psi^{(n)}_{i}(\mathbf{x},t),
	\label{fredholm_eqn}
\end{equation} 
where $\mathbf{W}(\mathbf{x})$ is a positive-definite Hermitian matrix accounting for the weights of each variable as defined in the (\ref{inner_product}). In  (\ref{fredholm_eqn}), $\lambda^{(n)}$ and $\Psi^{(n)}_{i}(\mathbf{x},t)$ correspond to the $n^{th}$ eigenvalue and the $i^{th}$ component of the $n^{th}$ eigenmode. The kernel $R_{ij}(\mathbf{x}, \mathbf{x}',t, t')$ which is the two-point two-time correlation tensor, is defined as follows:
\begin{equation}
	R_{ij}(\mathbf{x}, \mathbf{x}',t, t') = \langle u'_{i}(\mathbf{x},t) u'_{j}(\mathbf{x}',t') \rangle, \ \ \ \ i, j = 1, 2, 3, 
\end{equation}
\begin{equation}
	R_{i4}(\mathbf{x}, \mathbf{x}',t, t') = \langle u'_{i}(\mathbf{x},t) \rho'(\mathbf{x}',t') \rangle, \ \ \ \ i = 1, 2, 3,
\end{equation}
\begin{equation}
	R_{4j}(\mathbf{x}, \mathbf{x}',t, t') = \langle \rho'(\mathbf{x},t) u'_{j}(\mathbf{x}',t') \rangle, \ \ \ \ j = 1, 2, 3, 
\end{equation}
\begin{equation}
	R_{44}(\mathbf{x}, \mathbf{x}',t, t') = \langle \rho'(\mathbf{x},t) \rho'(\mathbf{x}',t') \rangle. \ \ \ \  
\end{equation}
For statistically stationary flows, such as the turbulent stratified wake in the present case, the kernel $R_{ij}(\mathbf{x}, \mathbf{x}',t, t')$ is only a function of time difference $\tau = t-t'$, $\mathbf{x}$, and $\mathbf{x}'$. Furthermore, it can be Fourier-transformed in the temporal direction as follows:
\begin{equation}
	R_{ij}(\mathbf{x}, \mathbf{x}',\tau) = \int_{-\infty}^{\infty} S_{ij}(\mathbf{x}, \mathbf{x'}, f)e^{i2\pi f\tau}\diff 
	f, 
	\label{fourier_trans}
\end{equation}
where $S_{ij}(\mathbf{x},\mathbf{x}', f)$ is the Fourier transform of the kernel $R_{ij}(\mathbf{x}, \mathbf{x}',\tau)$. Using (\ref{fourier_trans}), the Fredholm-type eigenvalue problem in (\ref{fredholm_eqn}) can be transformed into an equivalent eigenvalue problem which is solved at each frequency $f$, following \cite{towne_spectral_2018},
\begin{equation}
	\int_{\Omega}S_{ij}(\mathbf{x},\mathbf{x}',f)\mathbf{W}(\mathbf{x}')\Phi^{(n)}_{j}(\mathbf{x}',f)\diff\mathbf{x}' = \lambda^{(n)}(f) \, \Phi^{(n)}_{i}(\mathbf{x},f),
\end{equation}
where $\lambda^{(n)}(f)$ are the eigenvalues at $f$ and $\Phi^{(n)}_{i}(\mathbf{x},f) = \Psi^{(n)}_{i}(\mathbf{x},t)e^{-i2\pi f t}$ are the modified eigenmodes. The eigenvalues are ordered such that $\lambda^{(1)}(f) \geq \lambda^{(2)}(f) \geq . \ . \ .\ \geq \lambda^{(n)}(f) $. The sum over all the eigenvalues at frequency $f$ equates to two-times the total fluctuation energy content, i.e., $\langle u'_iu'_i \rangle$ $+$ $\frac{g^{2}}{\rho^{2}_{o}N^{2}}\langle \rho'\rho'\rangle$ at that frequency. The obtained eigenmodes in the frequency space are spatially orthogonal to each other such that:
\begin{equation}
	\int_{\Omega} \boldsymbol{\Phi}^{*(n)}(\mathbf{x},f)\mathbf{W}(\mathbf{x})\boldsymbol{\Phi}^{(m)}(\mathbf{x},f)\diff \mathbf{x} = \delta_{mn},
	\label{orthonormal}
\end{equation}
where $\delta_{mn}$ is the Dirac-delta function.
\subsection{Numerical implementation of SPOD for current work}\label{subsection_numerical_spod}


In this work, we primarily present results from SPOD on two-dimensional planes at various $x/D$ $-$ ranging from $x/D = 10$ to $100$ $-$ sampled at a spacing of approximately $5D$. The domain of $10\leq x/D\leq 100$ spans: (i) $5 \leq Nt_2 \leq50$ for $\Fro = 2$ and (ii) $1 \leq Nt_{10} \leq 10$ for $\Fro = 10$ in terms of buoyancy time. In the radial direction, the SPOD domain spans $0 \leq r/D \leq 10$, resulting in a total of $N^{SPOD}_r = 333$ points. In the azimuthal direction, the number of grid points $N_{\theta} = 256$. 

For numerical implementation, the mean-subtracted data, consisting of $N$ temporal snapshots, is divided into $N_{blk}$ blocks with an overlap of $N_{ovlp}$ snapshots. Each block contains $N_{freq}$ entries: $\mathbf{Q} = [\mathbf{q}^{(1)}, \mathbf{q}^{(2)}, \mathbf{q}^{(3)}, \cdots \mathbf{q}^{(N_{freq})}]$. Here, $\mathbf{q}^{(i)} = [\mathbf{u'}^{(i)}, \rho'^{(i)}]^{T}$ where $\mathbf{u'}$ and $\rho'$ are velocity and density fluctuations, respectively. Thereafter, discrete Fourier transform (DFT) of each block is performed in the temporal direction and the ensemble of $N_{blk}$ Fourier realizations of any given frequency, let us say $f$, is collected as $\mathbf{\hat{Q}}_f = [\mathbf{q}^{(1)(f)}, \mathbf{q}^{(2)(f)}, \mathbf{q}^{(3)(f)}, \cdots \mathbf{q}^{(N_{blk})(f)}]$. Once, $\mathbf{\hat{Q}}_{f}$ is obtained, SPOD eigenvalues and eigenvectors corresponding to $f$ are given by the following eigenvalue decomposition:

\begin{equation}
	\hat{\mathbf{Q}}^{*}_{f}\mathbf{W}\hat{\mathbf{Q}}_{f} \boldsymbol{\Gamma}_{f} = \boldsymbol{\Gamma}_{f}\boldsymbol{\Lambda}_{f},
	\label{spod_discrete}
\end{equation}

where $\boldsymbol{\Lambda}_{f} = diag\Big(\lambda^{(1)}_{f},\lambda^{(1)}_{f}, \cdots \lambda^{(N_{blk})}_{f}\Big)$ is a diagonal matrix containing eigenvalues ranked in the decreasing order of energy content from $i = 1$ to $N_{blk}$. The corresponding spatial eigenmodes $\hat{\boldsymbol{\Phi}}_{f}$ can be obtained as $\hat{\boldsymbol{\Phi}}_{f} = \hat{\mathbf{Q}}_{f}\boldsymbol{\Gamma}_{f}\boldsymbol{\Lambda}_{f}^{-1/2}$. In (\ref{spod_discrete}), $\mathbf{W}$ is a diagonal matrix of size $4N^{SPOD}_{r}N_{\theta}$, containing the numerical quadrature weights multiplied by coefficients required to form the energy quantities given in (\ref{inner_product}). 

The parameters for SPOD are set as follows: (i) total number of snapshots $N=7168$ with consecutive snapshots separated by $\Delta tD/U_{\infty} \approx 0.09$ and $0.104$ for $\Fro = 2$ and $10$, respectively, (ii) number of frequencies $N_{freq} = 512$, and (iii) overlap between blocks $N_{ovlp} = 256$, resulting in total of $N_{blk} = \frac{N-N_{ovlp}}{N_{freq} - N_{ovlp}} = 27$ SPOD modes at each frequency. Interested readers are referred to  \cite{towne_spectral_2018} and \cite{schmidt2020guide} for more details on the theoretical aspects and numerical implementation of SPOD.

Most of the results are obtained from SPOD analyses at constant $x/D$ planes with modes maximizing the two-times sum of TKE and TPE. However, for some results, we perform additional SPOD analyses. For example, to illustrate the streamwise variation of a certain leading-order SPOD mode in \secref{section_slice_spod_eigenmodes}, we perform SPOD analysis on fluctuating velocity and density fields at the center-vertical plane ($y=0$ plane) with reduced number of snapshots $N=5376$ and half-resolution in vertical and streamwise directions. $N_{freq}$ and $N_{ovlp}$ are kept the same as SPOD on fixed $x/D$ planes. The spatial resolution and $N$ are reduced to avoid memory limitations since large matrices with complex double precision have to be stored in the intermediate steps of SPOD. Also in \secref{section_VSmode}, we present results from SPOD analyses of the $\Fro =2$ wake (at constant $x/D$ planes) with (i) density fluctuations replaced by pressure fluctuations and (ii) norm defined such as to maximize the sum of $\langle p'p'\rangle$ and $\langle u'_iu'_i \rangle$. $N,N_{freq}$, and $N_{blk}$ are kept the same as in the previous paragraph. The motivation behind performing this additional set of SPOD analyses is explained in \secref{section_VSmode}. 

\section{Flow visualizations}\label{section_flow_viz}
Three-dimensional visualizations of the $Q$ criterion and planar views of the vorticity and velocity fields in this section provide a first look at the  vortical and unsteady IGW structure of the simulated wakes.
The structure of the steady (in a frame attached to the disk) lee wave field is not discussed in this paper. 
\begin{figure}
	\centering
	\includegraphics[width=\linewidth, keepaspectratio]{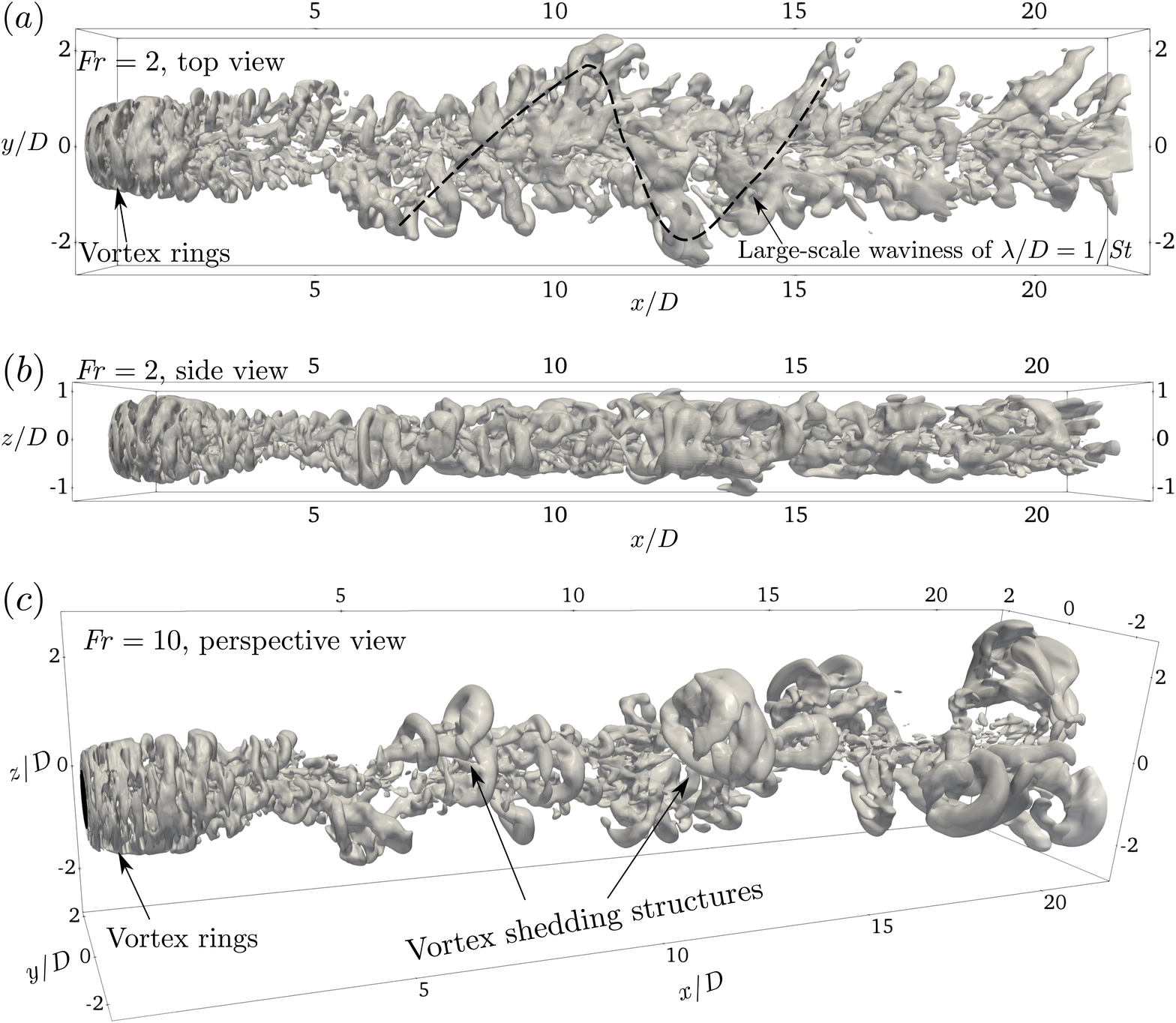}
	\caption{Isosurfaces of instantaneous $Q$ criterion at  $Q = 0.01$: (a,b) $\Fro = 2$ and (c) $\Fro = 10$. Streamwise domain is limited to $0 < x/D < 20$ for clarity.}
	\label{fig:qcri_fr2fr10}
\end{figure}
To emphasize  the large-scale coherent structures, the instantaneous velocity fields have been filtered using a SciPy Gaussian low-pass filter (\textit {gaussian\_filter}) in all three directions with standard deviation $\sigma = 5$ before calculating the $Q$ criterion and vorticity fields. 

Figure \ref{fig:qcri_fr2fr10} shows that, in both wakes, circular vortex rings appear immediately downstream  of the disk.  At   $\Fro = 2$, the buoyancy induced anisotropy between horizontal and vertical directions  commences in the near wake. The  wake contracts in the vertical at $x/D \approx 5$ (visible in the side view given in figure \ref{fig:qcri_fr2fr10}(b)) owing to the oscillatory modulation by the lee wave. The top view  (figure \ref{fig:qcri_fr2fr10}(a)) shows a distinct large-scale waviness in the intermediate wake, shown by the dashed black line. Its approximate wavelength is $\lambda/D \approx 1/\Str_{VS}$, where $\Str_{VS}$ is the vortex shedding (VS) frequency.
Likewise, large-scale VS structures separated approximately by $\lambda/D \approx 1/\Str_{VS}$ can also be identified in the $\Fro = 10$ wake (figure \ref{fig:qcri_fr2fr10}(c)). The value of $\Str_{VS}$ and the spatial behavior of the VS mode  will be made precise formally using SPOD in the subsequent sections.

%


\begin{figure}
	\centering
	\includegraphics[width=\linewidth, keepaspectratio]{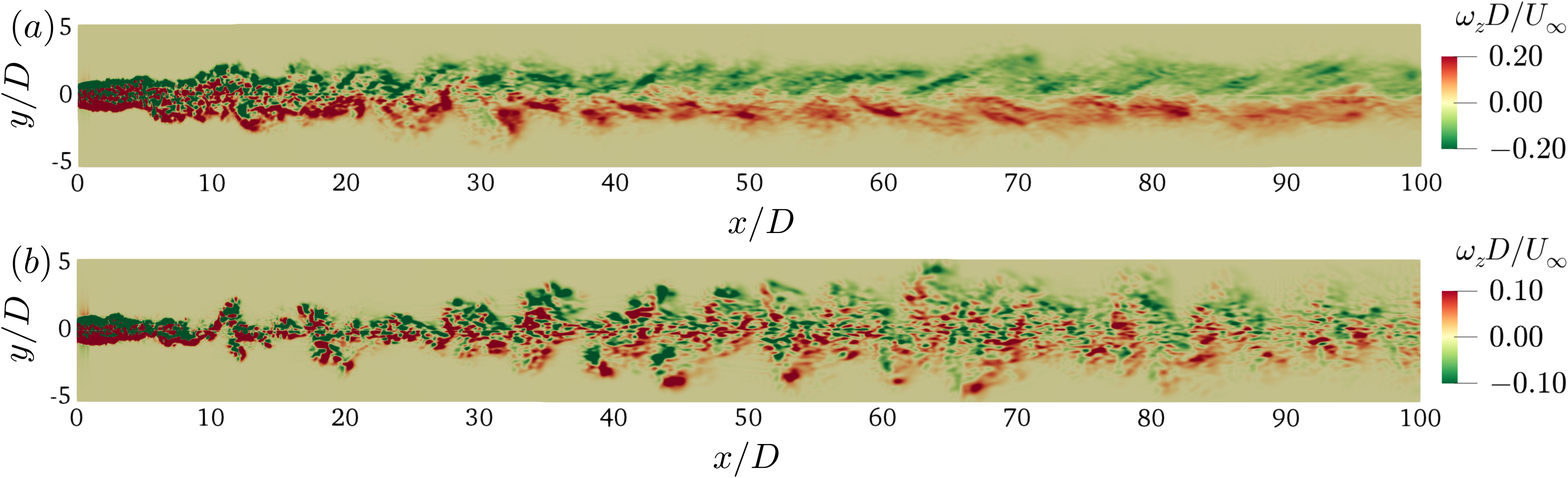}
	\caption{Instantaneous snapshot of vertical vorticity 
		at the central horizontal plane ($z = 0$): (a) $\Fro = 2$ and (b) $\Fro=10$. 
	}
	\label{fig:horiz_omgt_fr2fr10}
\end{figure}

Figure \ref{fig:horiz_omgt_fr2fr10} shows the instantaneous vertical vorticity ($\omega_{z}D/U_{\infty}$) on the central horizontal plane ($z = 0$)
for the $\Fro = 2$ (top) and $\Fro = 10$ (bottom) wakes. Similar to figure \ref{fig:qcri_fr2fr10}, $\omega_z$ is calculated using filtered velocity fields to emphasize large-scale features. In both wakes, the complex spatial distribution of  vorticity in the immediate downstream of the disk gives way to a well-defined coherent distribution of opposite signed vortices in the intermediate to late wakes. 
For the $\Fro = 2$ wake, spatial coherence is visible as early as $x/D \approx 20$. Beyond $x/D \approx 20$, the  regions of opposite-signed $\omega_{z}$ remain separated till the end of the domain. On closer inspection, a streamwise undulation of length $\lambda/D \approx 1/\Str_{VS}$ can be observed in figure \ref{fig:horiz_omgt_fr2fr10}(a). At this point, it is important to emphasize that the $\Fro = 2$ wake remains actively turbulent throughout the computational domain as demonstrated by CS2020 through spectra  and visualizations of the turbulent dissipation rate. 
From $x/D \approx 40$ onward, the $\Fro =2$ wake resides in the strongly stratified turbulent (SST) regime. Different regimes of stratified turbulence are discussed briefly in \secref{section_introduction}. 
The strong signature of coherence in the $\Fro = 2$ wake is not a consequence of the transition into the weakly turbulent state of the Q2D regime noted in previous works, e.g. by \cite{spedding_evolution_1997}. 

The $\Fro = 10$ wake also shows a distinct wavy motion with non-dimensional wavelength  $\approx 1/\Str_{VS}$. However, the separation between the regions with  opposite signed vorticity  is not as well defined as in the $\Fro = 2$ wake. According to CS2020, 
the $\Fro = 10$ wake stays in the weakly stratified regime (WST) from $x/D \approx 10$ to $50$ and thereafter stays in the intermediately stratified regime (IST) till the end of the domain.

\begin{figure}
	\centering
	\includegraphics[width=\linewidth, keepaspectratio]{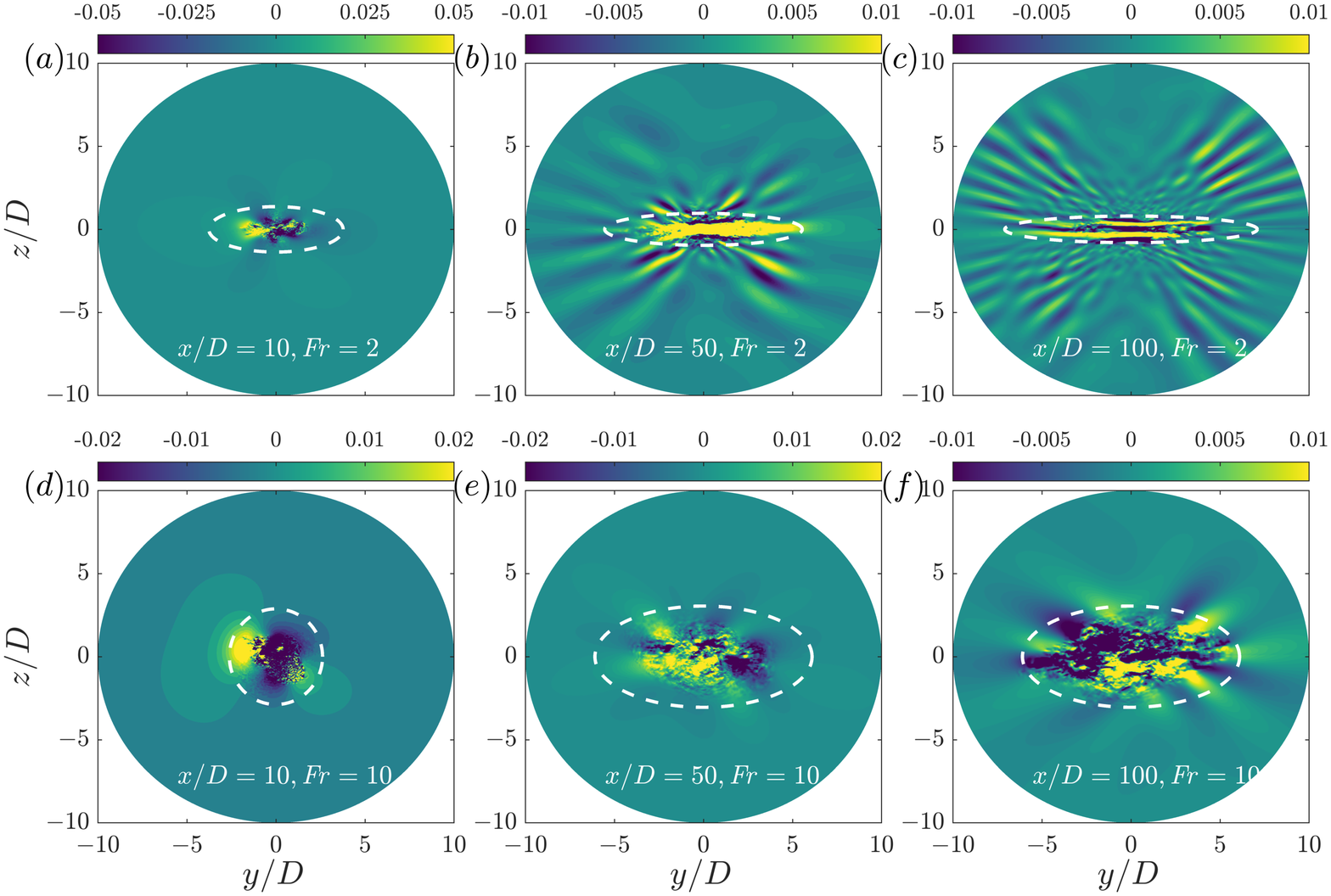}
	\caption{Instantaneous snapshots of the fluctuating spanwise velocity $u'_{y}/U_{\infty}$ shown for $\Fro = 2$ (top row) and $\Fro = 10$ (bottom row): (a,d) at $x/D = 10$, (b,e) at $x/D = 50$, and (c,f) at $x/D = 100$. Dashed close curve in white shows wake core.}
	\label{fig:slice_fr2fr10}
\end{figure}

To  conclude
this section,  instantaneous snapshots of fluctuating spanwise velocity ($u'_{y}/U_{\infty}$) are shown in figure \ref{fig:slice_fr2fr10} at locations in the near, intermediate and far wake 
at  $\Fro  = 2$ (top row)   and $\Fro = 10$ (bottom row).  An ellipse with major and minor axes equal to $2L_{Hk}$ and $2L_{Vk}$, where $L_{Hk}$ and $L_{Vk}$ are the  TKE-based wake widths in horizontal and vertical directions, respectively,  is also shown. $L_{Hk}$ is defined by $\mathrm{TKE}(x,y=L_{Hk},z=0) =\mathrm{TKE}(x,r=0)/2$ and $L_{Vk}$ by $\mathrm{TKE}(x,y=0,z=L_{Vk}) =\mathrm{TKE}(x,r=0)/2$. Here, $r=0$ denotes the disk centerline. It is worth noting that using the sum of TKE and TPE to define the wake widths (not shown here) result in values similar to $L_{Hk}$ and $L_{Vk}$ for both $\Fro = 2$ and $\Fro = 10$ wakes. Following CS2020, we use the TKE-based definitions in the rest of the results and discussions.
This ellipse, based on $L_{Hk}$ and $L_{Vk}$, is used to approximately demarcate the wake core from the outer wake. 
In subsequent sections, 
this definition of the wake core will prove to be useful for the interpretation of some SPOD results.

At $\Fro = 2$, an appreciable effect of buoyancy is already present in the near wake as shown in   figure \ref{fig:slice_fr2fr10}(a) for  $x/D = 10$, which corresponds to $Nt_{2} = 5$ in buoyancy time units.
At the same streamwise location, the $\Fro = 10$ wake still has a circular cross-section with an imprint of the $m=1$ azimuthal mode which was found to be energetically important in the unstratified wake (\cite{nidhan_spectral_2020}). As both the wakes evolve  downstream, buoyancy has a progressively increasing effect on the the wake core as well as the surrounding outer wake. By $x/D = 50$, vertically flattened wake cores can be observed in figure \ref{fig:slice_fr2fr10}(b,d) for both the wakes, more so at $\Fro = 2$ than at  $\Fro = 10$. It is also worth noting that the wake core of $\Fro = 2$ consists of distinct layers by $x/D  = 50$. The $\Fro =2$ wake also shows a significant amount of IGW activity in the  outer region, i.e. outside the ellipse in figure \ref{fig:slice_fr2fr10}(b). 
Farther downstream at $x/D = 100$, the $u'_{y}$ field of $\Fro = 2$ (figure \ref{fig:slice_fr2fr10}(c)) shows IGWs occupying a significant portion of the outer wake with the wake core being further flattened 
and comprising an increased number of  horizontally oriented layers. The $\Fro = 10$ wake core also starts showing appreciable IGW activity in the ambient by $x/D = 100$ ($Nt_{10}= 10$), as shown in figure \ref{fig:slice_fr2fr10}(f). 

\section{Characteristics of SPOD eigenvalues and eigenspectra}\label{section_results_eigenvalues_eigenspectra}

We start the discussion of  SPOD modes by evaluating their overall contribution to fluctuation energy and by their eigenspectra. There are significant effects of buoyancy as elaborated below.

\subsection{Cumulative modal contribution to  fluctuation energy} \label{subsection_cumulative_energy_plot}

\begin{figure}
	\centering
	\includegraphics[width=\linewidth, keepaspectratio]{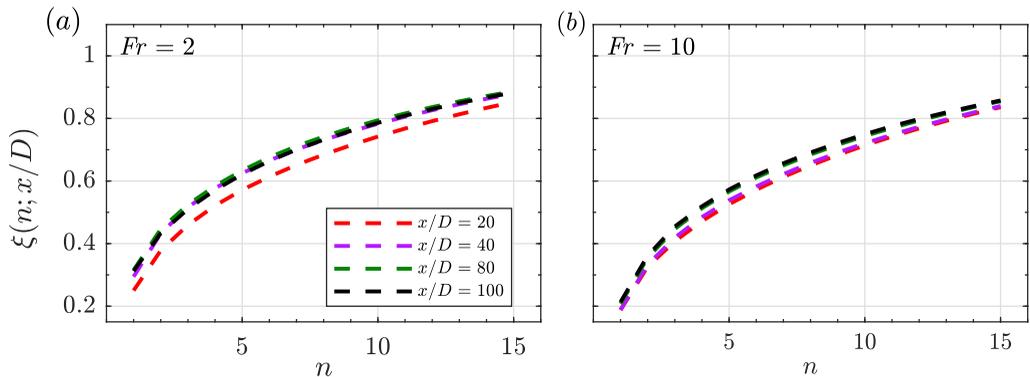}
	\caption{Variation of the cumulative energy, $\xi(n)$, as a function of modal index $n$ for (a) $\Fro= 2$ and (b) $\Fro = 10$ wakes, shown till $n = 15$ SPOD modes. 
	} 
	\label{fig:modal_indx_energy_fr2fr10}
\end{figure}

Figure \ref{fig:modal_indx_energy_fr2fr10} shows the variation of  cumulative energy ($\xi(n)$) as a function of SPOD modal index ($n$) at 
four downstream locations: $x/D = 20, 40, 80,$ and $100$.
To calculate $\xi(n)$, the energy across all resolved frequencies $\Str$ at each modal index up to $n$ is summed and normalized by the total energy as follows:
\begin{equation}
	\xi(n;x/D) = \frac{\sum\limits_{i=1}^{n}\sum_{St} \lambda^{(i)}(f;x/D)}{\sum\limits_{i=1}^{N_{blk}}\sum_{St} \lambda^{(i)}(f;x/D)},
	\label{eigenpectrum_int_cummul_energy}
\end{equation}
where $N_{blk}$ is the total number of SPOD modes at a given $\Str$.
Comparison among the various $x/D$ curves shows that the energy captured by leading SPOD modes in both wakes increases with downstream distance.
This behavior is in contrast to the unstratified wake where the relative importance of the dominant SPOD modes decreases with increasing $x/D$  \citep{nidhan_spectral_2020}.  
Although both stratified wakes exhibit an increasing dominance of the leading modes as $x/D$ increases, there is a
quantitative difference in that the jump of modal energy fraction  from its $x/D = 20$ value is   larger for the  the $\Fro = 2$  wake relative to the $\Fro = 10$ wake. 



As discussed in the introduction, CS2020
found that  the $\Fro =2$ wake traversed the WST, IST and SST regimes  during its streamwise evolution and the $\Fro = 10$ wake accessed only the WST and IST regimes. Readers are referred to \secref{section_introduction} for an introduction to these regimes in stratified turbulence. These transitions also appear in the the  evolution of the modal energy content $\xi(n;x/D)$.  For example, the  $\Fro = 2$ wake in  figure \ref{fig:modal_indx_energy_fr2fr10}(a) shows a transition at $x/D \approx 40$ whereby  the  $\xi(n)$ curves  for $x/D \geq 40$ collapse onto a single profile.  This result is consistent with CS2020 who  find that  $x/D \approx 40$ ($Nt_{2} \approx 20)$ is the location where the $\Fro = 2$ wake  wake transitions from IST to SST.
The $\Fro = 10$ wake was found by CS2020
to stay in the WST regime till $x/D \approx 50$ ($Nt \approx 5$)  and thereafter transitioned to the IST regime.
For the $\Fro = 10$ wake in figure \ref{fig:modal_indx_energy_fr2fr10}(b), the $\xi(n)$ curves collapse separately, i.e. there is one curve showing collapse between  $x/D = 20$ and 40  which lie in the WST regime, and there is another  showing collapse between $x/D = 80$ and $100$ which lie in the IST regime.
Plots of $\xi(n)$ at other values of  $x/D$ (not shown here) confirm that locations with $x/D \leq 50$ collapse on the $x/D = 20,40$ curve and locations with $x/D \geq 80$ collapse on the $x/D = 80,100$ curve.

\begin{figure}
	\centering
	\includegraphics[width=\linewidth, keepaspectratio]{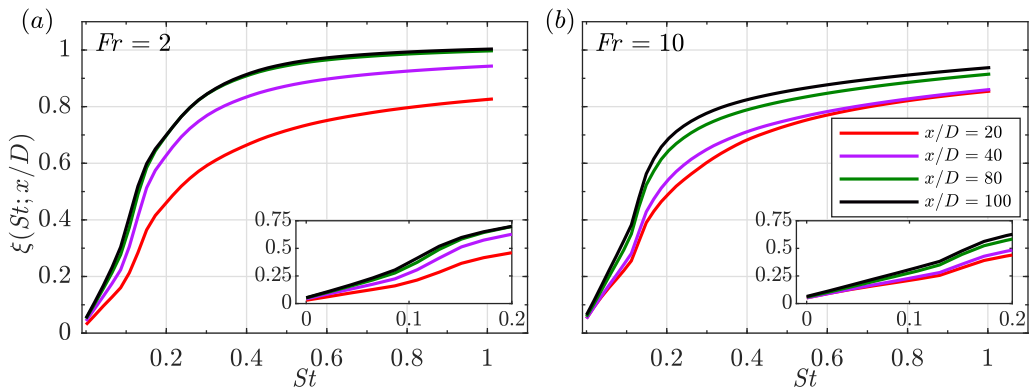}
	\caption{Variation of the cumulative fraction of energy, $\xi(\Str)$, as a function of $\Str$ for (a) $\Fro= 2$ and (b) $\Fro = 10$. The plots are shown for $0 \leq  \Str \leq 1 $ for both cases. Inset plots show zoomed-in variation of $\xi(\Str)$ for $0 \leq \Str \leq 0.2$.}
	\label{fig:freq_energy_fr2fr10}
\end{figure}

The energy summed over frequencies instead of modal indices is now examined.
Figure \ref{fig:freq_energy_fr2fr10}  shows the variation of $\xi(\Str)$
calculated as follows:
\begin{equation}
	\xi(\Str; x/D) = \frac{\sum\limits_{f = -St}^{St}\sum\limits_{i=1}^{N_{blk}} \lambda^{(i)}(f;x/D)}{\sum\limits_{i=1}^{N_{blk}}\sum_{St} \lambda^{(i)}(f;x/D)}.
	\label{mode_int_cummul_energy}
\end{equation}
Figure \ref{fig:freq_energy_fr2fr10} shows that  $\xi(\Str)$ increases for low-$\Str$ modes with increasing $x/D$ in both wakes, which is a trend
also seen for $\xi(n)$.
This is yet another indication of the increasing importance of the coherent  modes as  buoyancy effects come into play in these stratified wakes.  Besides, for both wakes, $\xi(\Str)$ increases steeply between $\Str = 0.1$ and $ 0.2$ at all downstream locations. The reason behind this sharp increase will be discussed shortly.  Another observation of interest is that almost all the fluctuation energy  at large $x/D$ is captured by the modes with $\Str < 1$ in both wakes.  


From $x/D = 20$ to $40$, there is a large jump in  $\xi(\Str)$ for the $\Fro = 2$ wake in figure \ref{fig:freq_energy_fr2fr10}(a).
As mentioned previously, $x/D = 40$ also marks the arrival of the $\Fro = 2$ wake into the SST regime.  Also, the $\xi(\Str)$ curves collapse for locations $x/D = 80$ and $100$. On analyzing other streamwise locations (not shown here), we find that the $\xi(\Str)$ curves for $x/D \geq 70$ collapse together similar to  the previously shown $\xi(n)$ curves of the $\Fro = 2$ wake. One difference is that the collapse of $\xi(n)$ commences closer to the body at $x/D \approx 40$.  

Contrary to the $\Fro = 2$ wake where the
change in $\xi(\Str)$  from $x/D = 20$ to $x/D = 40$ was large, the corresponding change  for the $\Fro = 10$ wake (figure \ref{fig:freq_energy_fr2fr10}(b))  is small and consistent with an absence of regime change.
However, the  $\Fro = 10$ wake exhibits a significant jump of $\xi(\Str)$  between $x/D = 40$ and $80$, which lie in the WST and IST regime, respectively. 

	To summarize, figures \ref{fig:modal_indx_energy_fr2fr10} and \ref{fig:freq_energy_fr2fr10} have the following implications. First, the relative importance of the leading SPOD modes increases with $x/D$ for the stratified wakes, which is in stark contrast to their  behavior in the unstratified wake (\cite{nidhan_spectral_2020}). Second, the trend of increasing dominance of leading SPOD modes is more pronounced for the strongly stratified wake of $\Fro = 2$ as compared to $\Fro = 10$. Third, transitions between WST, IST and SST regimes discussed by CS2020 for the turbulence statistics are also qualitatively reflected in the energetics of SPOD modes too.

\subsection{SPOD eigenspectra of $\Fro = 2$ and $10$ wakes}\label{subsection_slice_spod_eigenspectra}

\begin{figure}
	\centering
	\includegraphics[width= 0.8\linewidth]{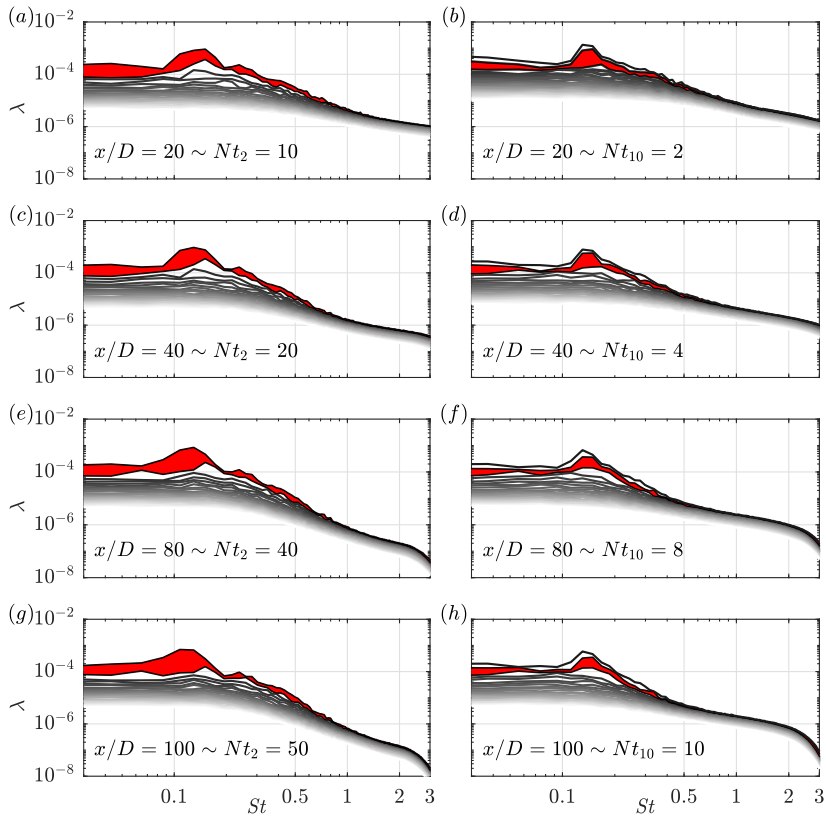}
	\caption{SPOD eigenspectra of 25 most energetic modes, $\lambda^{(1)}$ to $\lambda^{(25)}$, for the $\Fro = 2$ (left column) and $\Fro = 10$ (right column) wakes at four streamwise locations: (a,b) $x/D = 20$, (c,d) $x/D = 40$, (e,f) $x/D =80$, and (g,h) $x/D = 100$. Dark to light shade corresponds to increasing model index $i$ in $\lambda^{(i)}$.}
	\label{fig:fr2fr10_eigenspectrum}
\end{figure}

Figure \ref{fig:fr2fr10_eigenspectrum} shows the SPOD eigenspectra of the $\Fro = 2$ (left column) and $\Fro = 10$ (right column) wakes at  various
downstream locations.
The spectrum of the leading SPOD mode ($\lambda^{(1)}$) shows a distinct spectral peak in the vicinity of $\Str = 0.13-0.15$ at all locations and for both wakes. This pronounced peak is the reason why there was a sharp increase of $\xi(\Str)$  within $0.1 < \Str < 0.2$ for both wakes in figure \ref{fig:freq_energy_fr2fr10}. 

In the $\Fro = 10$ wake,  the $\lambda^{(1)}$ eigenspectrum at all locations has a distinct  peak at $\Str \approx 0.13$,
which  is very close to the vortex shedding (VS) frequency of the unstratified wake ($\Str = 0.135$) at the same Reynolds number (\cite{nidhan_spectral_2020}). SPOD eigenspectra 
at $x/D < 2$ (not presented here) show that this spectral peak has its origin near the wake generator and corresponds  to vortex shedding  in the $\Fro = 10$ wake. 

Unlike the $\Fro = 10$ wake, the spectral peak in $\lambda^{(1)}$ for the $\Fro = 2$ wake shifts slightly from $\Str \approx 0.15$ at $x/D = 20$ to $\Str \approx 0.13$ at $x/D = 40$ and onward.
At the far wake location of $x/D = 80$, the peak in the $\lambda^{(1)}$ eigenspectrum broadens to reach $\Str \approx 0.11$. Near-body SPOD eigenspectra (not shown here) for the $\Fro = 2$ wake show a prominent peak at $\Str \approx 0.15$ (slightly larger relative to the unstratified and $\Fro = 10$ wakes)  just downstream of the recirculation zone at $x/D \approx 2$. Furthermore the pressure spectrum (not presented here) in the immediate proximity of the disk, at $x/D = 0.5$ and $r/D=0.5$, also peaks at $\Str \approx 0.15$, indicating that this frequency corresponds to the VS mechanism for the $\Fro = 2$ wake. 
The shift in the spectral peak towards lower $\Str$ at later $x/D$ is consistent
with the  sphere-wake study of \cite{spedding_streamwise_2002} who report a gradual reduction in the dominant  wake $\Str$ during $40 < Nt < 100$ (see figure 5 of  their paper).

In the $\Fro = 2$ wake,  there is a large gap (demarcated in red) between the $\lambda^{(1)}$ and $\lambda^{(2)}$ spectra for frequencies with $\Str < 0.2$.
Beyond $\Str \approx 0.2$, values of all $\lambda^{(i)}$ fall sharply. This large difference between $\lambda^{(1)}$ and $\lambda^{(2)}$ implies that the dynamics of the $\Fro=2$ wake is  low-rank, i.e.  it is  dominated by the  leading SPOD mode. The sharp drop-off in energy at higher $\Str$ points to the dominance of low-frequency  energetic structures with $\Str $ in $[0, 0.2]$, specifically around the VS frequency.

In terms of low-rank behavior, the $\Fro = 10$ wake shows a peculiar difference from the $\Fro = 2$ wake.  Although the gap between $\lambda^{(1)}$ and $\lambda^{(2)}$  is significantly less compared to that for $\Fro =2$, 
there is a significant gap between $\lambda^{(2)}$ and $\lambda^{(3)}$ around the VS frequency $\Str \approx 0.13$, shown in red in the right column of figure \ref{fig:fr2fr10_eigenspectrum}. Furthermore, the variation of $\lambda^{(2)}$ with $\Str$ is very similar to that of $\lambda^{(1)}$. On further investigation, we find that the SPOD eigenmodes of $\lambda^{(1)}$ and $\lambda^{(2)}$ at the VS frequency have similar spatial structure, but with a rotation in their orientation. We hypothesize that $\lambda^{(1)}$ and $\lambda^{(2)}$ modes at the VS frequency are the manifestation of $m=1$ and $m=-1$ azimuthal modes in the weakly stratified $\Fro = 10$ wake.


\begin{figure}
	\centering
	\includegraphics[width=\linewidth, keepaspectratio]{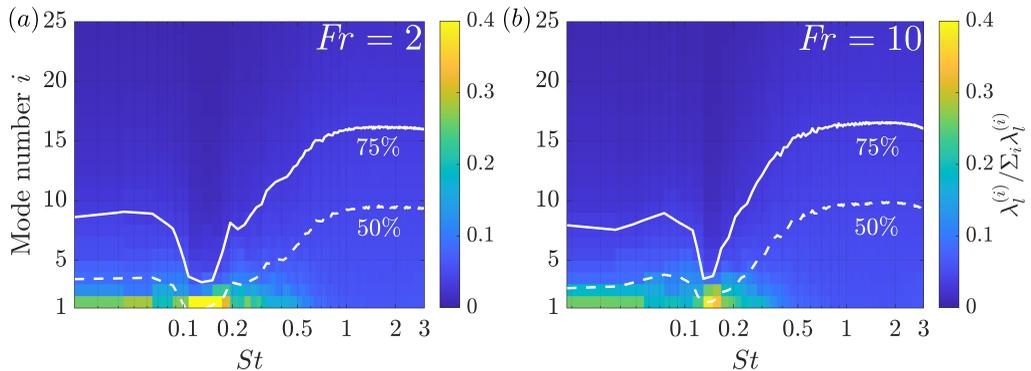}
	\caption{Fraction of energy (at a given $\Str$) accounted by each SPOD mode as a function of frequency at $x/D = 50$ for: (a) $\Fro = 2$ and (b) $\Fro = 10$ wakes. The solid and dashed white lines indicate the number of SPOD modes required to retain 75\% and 50\% of the total fluctuation energy respectively at each frequency.} 
	\label{fig:st_i_fr2fr10}
\end{figure}

Figure \ref{fig:st_i_fr2fr10} shows the fraction of energy in each SPOD mode as a function of $\Str$ for both wakes at a representative location of $x/D = 50$. In both wakes, the leading SPOD mode at the VS frequency capture at least $40\%$ of the total energy contained in the VS frequency. This also holds true for the near ($x/D = 10$) and far ($x/D = 100$)  locations in both wakes (not discussed here for brevity). Also, in both, wakes, less than 5 SPOD modes are required to capture $75\%$ of the total energy in the vicinity of the VS frequency, as indicated by the solid white line in figure \ref{fig:st_i_fr2fr10}.

	The key takeaway from figure \ref{fig:fr2fr10_eigenspectrum} and figure \ref{fig:st_i_fr2fr10} is twofold: (i) the VS frequency is the leading contributor to the fluctuating energy content of both $\Fro = 2$ and $10$ wakes and (ii) its dynamics are primarily governed by a few leading SPOD modes. Previous experimental studies of \cite{chomaz_structure_1993} and \cite{lin_turbulent_1992} have showed the existence of the VS mode in the near wake at moderate stratification using hot-wire measurements (at few select locations) and shadowgraph techniques. The present SPOD analysis enables us to establish the dominance of the VS mode
	in 
	stratified wakes 
	from near the body to 100 body diameters downstream by providing an ordered set of $\lambda^{(i)}$ eigenvalues for different $\Str$.

\section{The energetics of the vortex shedding (VS) mode}\label{section_VSmode}

A comparison between SPOD eigenspectra of the stratified wakes (figure \ref{fig:fr2fr10_eigenspectrum}) and the unstratified wake (figure 3 in \cite{nidhan_spectral_2020}) reveals that both types of wakes are dominated by vortex shedding which gives rise to a distinct spectral peak in the vicinity of $\Str \approx 0.13$. For the unstratified wake, besides the VS structure, which appears in the azimuthal mode $m=1$, a  double helix ($m=2$) mode with a peak at $\Str \rightarrow 0$ is also found to be energetically important \citep{johansson_far_2006,nidhan_spectral_2020}. In the stratified wake, as elaborated below, we find that the VS mode is persistent, is linked to unsteady internal gravity waves (IGWs), and is thereby responsible for the accumulation of fluctuation energy outside the wake core.

\begin{figure}
	\centering
	\includegraphics[width=\linewidth]{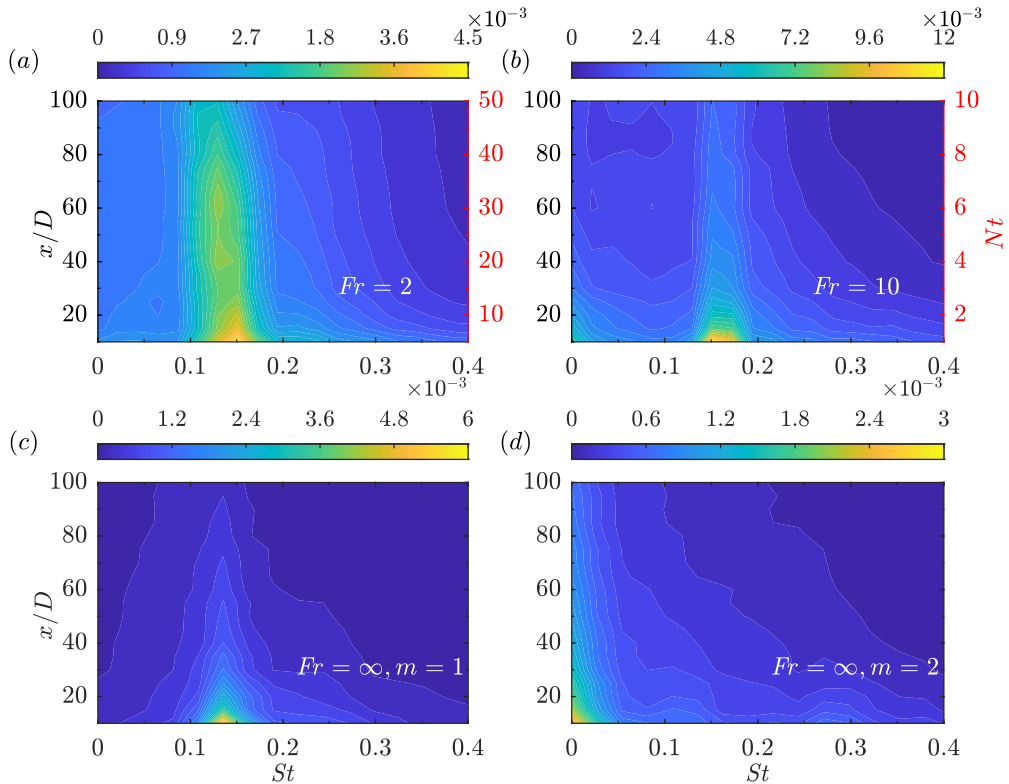}
	\caption{$x/D-\Str$ contour maps showing the variation of total energy in the leading 15 SPOD modes: (a) $\Fro = 2$, (b) $\Fro =10$, (c) $\Fro = \infty$, $m=1$ (vortex shedding) mode, and (d) $\Fro = \infty$, $m=2$ (double helix) mode. The contour level of each plot is set between zero and the maximum value of the energy over all $(x/D, \Str)$ pairs in that plot. Results for $\Fro = \infty$ wake are taken from \cite{nidhan_spectral_2020}.}
	\label{fig:active_frequencies_fr2fr10frinf_nblk15}
\end{figure}
\begin{figure}
	\centering
	\includegraphics[width=0.5\linewidth]{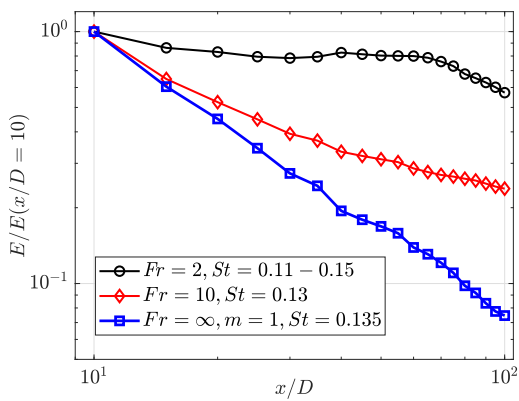}
	\caption{Evolution of the energy contained in leading 15 SPOD modes   at the vortex shedding frequency is shown for $\Fro = 2, 10$, and $\infty$ wakes. The energy is normalized by its value at $x/D = 10$. Energy across $0.11 \leq \Str \leq 0.15$ is summed for the $\Fro = 2$ wake since it has a broader spectral peak (see figure \ref{fig:fr2fr10_eigenspectrum}).
	}
	\label{fig:vs_mode_decay}
\end{figure}

Stratification qualitatively affects the streamwise evolution of the energy in different frequencies.
The evolution of the frequency-binned energy is shown for the stratified wakes in figure \ref{fig:active_frequencies_fr2fr10frinf_nblk15} (a)-(b). For the unstratified case ($\Fro = \infty$), the  azimuthal modes $m=1$ and $m=2$ are shown in figure \ref{fig:active_frequencies_fr2fr10frinf_nblk15}(c) and (d),  respectively.  For the stratified wakes in figures \ref{fig:active_frequencies_fr2fr10frinf_nblk15} (a)-(b), 
the spectral peak in the vicinity of $\Str \approx 0.13$ remains prominent 
for significant downstream distances, especially for $\Fro = 2$.
A somewhat wide band ($0.1 \leq \Str \leq 0.2$), centered around  $\Str \approx 0.13$ of the VS mode, is excited for the stratified wakes. Furthermore, this band persists into the far wake.  Even at $x/D = 100$,  this band has larger energy density than at other frequencies.  Such persistence in the energetic dominance of the VS mode (and neighboring frequencies) is absent in the unstratified $\Fro = \infty$ case where the energy at the two peaks of: (i) $\Str = 0.135$ in the $m=1$ mode (figure \ref{fig:active_frequencies_fr2fr10frinf_nblk15}(a)) and (ii) $\Str = 0$ in the $m=2$ mode (figure \ref{fig:active_frequencies_fr2fr10frinf_nblk15}(b)) declines sharply with increasing $x/D$.

	Figure \ref{fig:vs_mode_decay} shows the streamwise evolution of energy in the leading 15 SPOD modes and in a frequency band around the VS frequency.
	The energy in the $\Fro = 2$ wake remains almost constant till $x/D = 60$ and starts decaying slowly thereafter. On the other hand, the  $\Fro = 10$ wake shows an initial  decay in the VS mode energy 
	which  closely follows that of the $\Fro = \infty$ wake till $x/D = 20$. Subsequently, buoyancy effects set in for the $\Fro = 10$ wake  to slow down the energy decay.   


\begin{figure}
	\centering
	\includegraphics[width=\linewidth]{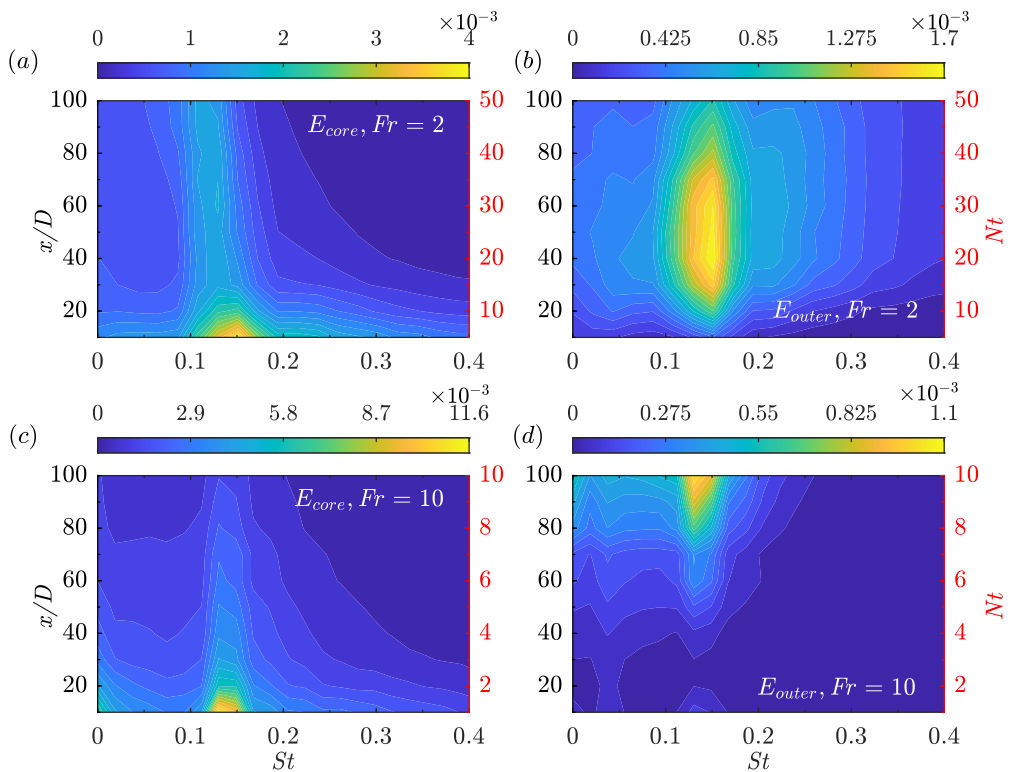}
	\caption{ Energy partition between core and outer wake: (a) wake core of $\Fro=2$, (b) outer wake of $\Fro =2$, (c) wake core of $\Fro =10$, and (d) outer wake of $\Fro = 10$. The first 15 SPOD modes are considered and the contour level of each plot is set between zero and the maximum value of the energy over all $(x/D, \Str)$ pairs in that plot.}
	\label{fig:wake_wave_partition_fr2_fr10}
\end{figure}

To investigate the reason behind the downstream persistence of the VS spectral peak in  stratified wakes,  the total energy in the  leading 15 SPOD modes is partitioned into two components: (i) energy of the wake core, $E_{core}$ and (ii) energy of the outer wake, $E_{outer}$. The energy in each of the regions is calculated as:
\begin{equation}
	E_{core}(x/D,\Str) = \sum\limits_{n=1}^{15} \ \int\limits_{A \in \Omega} \lambda^{(n)}(x/D,\Str)  \Phi^{*(n)}_{i}(x/D,\Str)\Phi^{(n)}_{i}(x/D,\Str)\diff A,
	\label{eq:energy_wake_core}
\end{equation}
\begin{equation}
	E_{outer}(x/D,\Str) = \sum\limits_{n=1}^{15} \ \int\limits_{A \in \mathcal{H} - \Omega} \lambda^{(n)}(x/D,\Str) \Phi^{*(n)}_{i}(x/D,\Str)\Phi^{(n)}_{i}(x/D,\Str)\diff A,
	\label{eq:energy_wake_ambient}
\end{equation}
where $\Omega$ denotes the wake core at a given $x/D$, as defined in \secref{section_flow_viz}. $\mathcal{H}$ denotes the area of  the circular cross-section bounded by $0 \leq r/D \leq 10$ at a given $x/D$. Here, $\Phi^{(n)}_{i}$ corresponds to the $n^{th}$ SPOD eigenmode for a given $x/D$ and $\Str$.

The energy in the wake core peaks around the VS-mode frequency, $\Str \approx 0.12-0.13$,  for both wakes (see figure \ref{fig:wake_wave_partition_fr2_fr10}(a,c)). With increasing $x/D$ (or $Nt$), the VS signature in the wake core decays for both wakes. 
The energetics of the outer wake is remarkably different.
$E_{outer}$, 
which starts off with a small value across all $\Str$ at $x/D \approx 10$ in the $\Fro = 2$  wake, develops a peak at $\Str \approx 0.15$ at $x/D \approx  20$. Note that this peak is the  same as the peak in the SPOD eigenspectrum for the entire wake  (figure \ref{fig:fr2fr10_eigenspectrum}(a)). Farther downstream, 
there is significant energy content in 
the outer wake for $x/D \approx 16 - 80$ ($Nt_{2} \approx  8 - 40$) with a spectral peak located at $\Str \approx 0.13-0.15$.
The spectral peak is broad,
i.e.  nearby frequencies with $0.1 \leq \Str \leq 0.2$ also have comparable energy levels. For the  $\Fro = 10$ wake, $E_{outer}$ picks up only beyond $x/D = 60$ ($Nt_{10} = 6$), and thereafter increases progressively in the vicinity of $\Str \approx 0.13$ till the end of the domain. We also find that the qualitative nature of the variation of energy in the outer wake and wake core  found in figure \ref{fig:wake_wave_partition_fr2_fr10} does not change when the number of modes over which energy is summed is decreased from 15 to 3 (not presented here for brevity).

	
	\begin{figure}
		\centering
		\includegraphics[width=\linewidth]{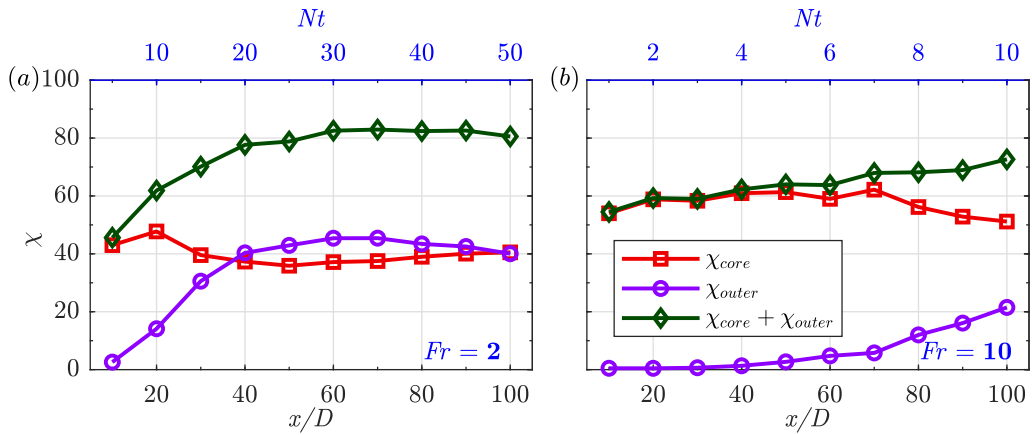}
		\caption{Streamwise variation of $\chi_{core}$, $\chi_{outer}$, and $\chi_{core} + \chi_{outer}$:  (a) $\Fro = 2$, (b) $ \Fro  =10$.}
		\label{fig:wake_wave_partition_fr2_fr10_nt}
	\end{figure}
	
	In figure \ref{fig:wake_wave_partition_fr2_fr10_nt}, we sum up the SPOD energies across $\Str \in [-0.4, 0.4]$ separately for the wake core and the outer wake and compute their  percentage  contribution to the entire area-integrated fluctuation energy as follows,
	\begin{equation}
		\chi_{core}(x/D) =  \frac{\sum\limits_{|St|=[0, 0.4]}E_{core}(\Str,x/D)}{E_{k}^{T}(x/D) + E_{\rho}^{T}(x/D)} \times 100,
	\end{equation}
	\begin{equation}
		\chi_{outer}(x/D) =  \frac{\sum\limits_{|St|=[0, 0.4]}E_{outer}(\Str,x/D)}{E_{k}^{T}(x/D) + E_{\rho}^{T}(x/D)} \times 100, 
	\end{equation}	
	where $E^{T}_{k}(x/D)$ and $E^{T}_{\rho}(x/D)$ are area-integrated TKE and TPE in the circular region of $0 \leq r/D \leq 10$ at the $x/D$ location under consideration.  The streamwise evolution of $\chi_{core}$ and $\chi_{outer}$ are  shown in  figure \ref{fig:wake_wave_partition_fr2_fr10_nt}.
	
		For the $\Fro = 2$ wake (figure \ref{fig:wake_wave_partition_fr2_fr10_nt}(a)), $\chi_{outer}$ increases monotonically until $x/D \approx 60$ followed by a slight decrease.
		At its peak, $\chi_{outer}$  constitutes up to $50\%$ of the total fluctuation energy, becoming even larger than $\chi_{core}$. In the $\Fro = 10$ wake (figure \ref{fig:wake_wave_partition_fr2_fr10_nt}(b)), $\chi_{outer}$  remains negligible till $x/D = 60$, followed by a monotonic increase. The increase in the value of $\chi_{outer}$  is accompanied by a decrease in the  wake-core contribution.
	The percentage of total energy captured by the leading 15 SPOD modes and $|\Str| \in [0, 0.4]$, i.e., $\chi_{wake} + \chi_{outer}$ (shown in green), increases for both wakes from its initial value at $x/D = 10$. This reinforces a main finding of this work that stratified wakes display an increased coherence of fluctuation energy as they evolve downstream.
	
	Figure \ref{fig:wake_wave_partition_fr2_fr10} suggests that the unsteady IGWs in the outer wake radiate from the VS mode at intermediate to late $Nt$. Nevertheless, to further establish causation between the unsteady IGW emission and the VS mode, we perform additional SPOD analyses for the $\Fro = 2$ wake. In these analyses, we replace the fluctuating density field $\rho'$ with the fluctuating pressure field $p'$. These SPOD analyses are performed at $x/D = 10, 20, \cdots, 90,100$. At all locations, the eigenspectra obtained from these modified SPOD analyses show a prominent peak at the VS frequency with a large gap between $\lambda^{(1)}$ and $\lambda^{(2)}$ for $\Str < 0.2$, qualitatively akin to the left column of figure \ref{fig:fr2fr10_eigenspectrum}.
	
	Using $p'$ along with $u'_i$ enables us to reconstruct the pressure transport term in the radial direction, $\langle p'u'_r \rangle$,  which accounts for the energy transferred radially from the wake core to the IGW dominated outer wake region through pressure-work \citep{de_stadler_simulation_2012,rowe_internal_2020}. We reconstruct $\langle p'u'_r \rangle$ contours using leading 15 SPOD modes and frequencies in the range of (i) $\Str \in [0.1,0.2]$  and (ii) $\Str \in [0.1,0.3]$ as follows:
	
	\begin{equation}
		\langle p'u'_r \rangle(x;y,z) =  \sum\limits_{St }\sum\limits_{n=1}^{15}\lambda^{(n)}(x;\Str) \, \Phi_{u_r}^{(n)}(x;y,z,\Str) \, \Phi_{p}^{(n)*}(x;y,z,\Str),
		\label{eq:pur_recon}
	\end{equation}
	
	\begin{figure}
		\centering
		\includegraphics[width=\linewidth]{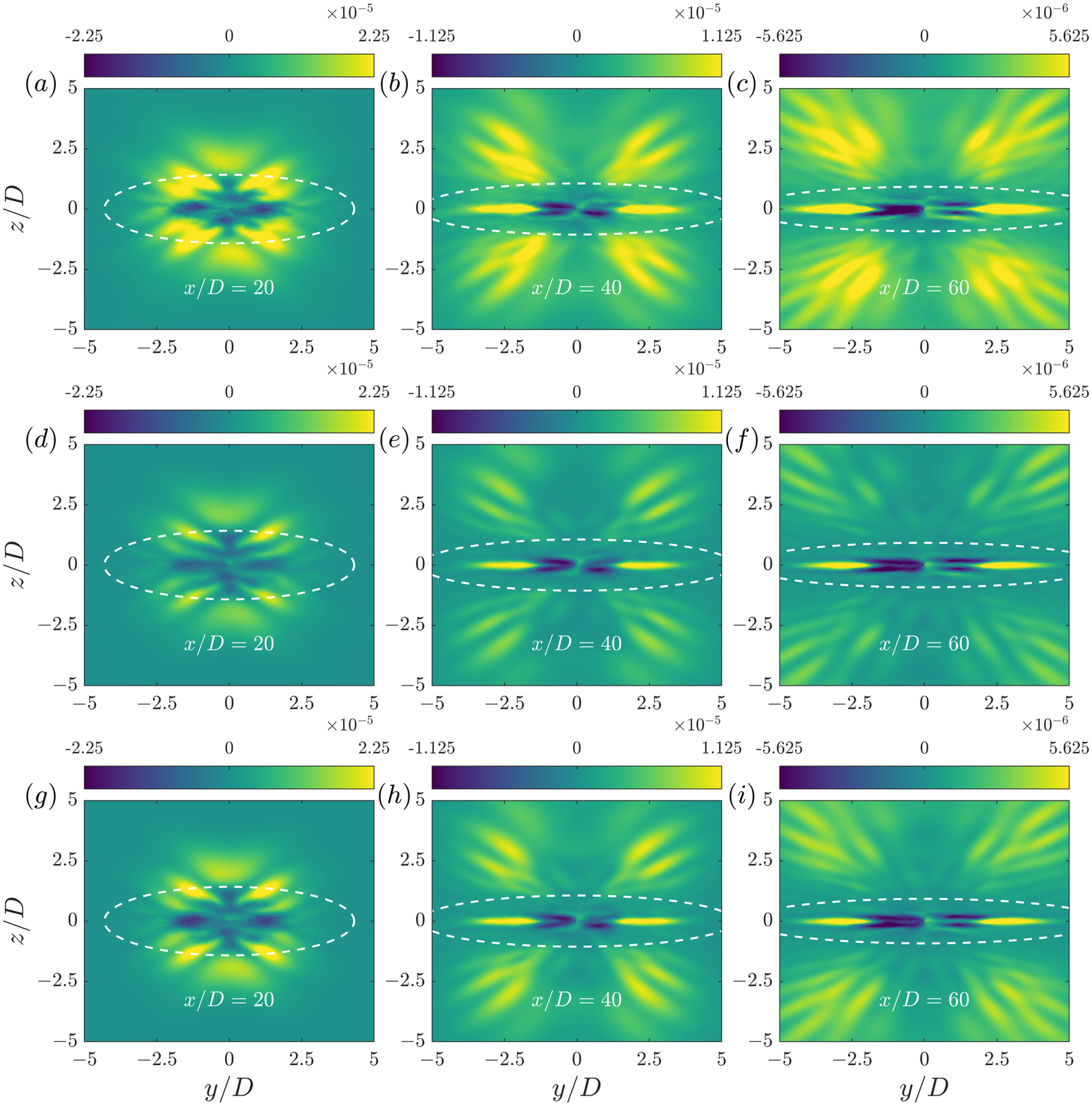}
		\caption{Contours of $\langle p'u'_r \rangle$ for the $\Fro = 2$ wake obtained from: (i) temporal averaging (top row), reconstruction using leading 15 SPOD modes and (ii) $\Str \in [0.1, 0.2]$ (middle row) and (iii) $\Str \in [0.1,0.3]$ (bottom row). Three streamwise locations $x/D = 20, 40,$ and $60$ are shown. Dashed closed curve in white shows wake core.}
		\label{fig:pur_reconstructed_nblk15_st01_02}
	\end{figure}
	
	Figure \ref{fig:pur_reconstructed_nblk15_st01_02} shows the actual (top row) and reconstructed (middle and bottom rows) $\langle p'u'_r \rangle$ at three streamwise locations $x/D = 20, 40, $ and $60$ for the $\Fro = 2$ wake. The actual $\langle p'u'_r\rangle$ shows a strong signature of IGW flux in the outer wake region at all three downstream locations in figure \ref{fig:pur_reconstructed_nblk15_st01_02}. We found that the nonlinear transport term was negligible outside the wake core (not shown here). Hence the primary source of the energy transfer to the outer wake is the pressure-work term due to the IGW radiation. The reconstructed $\langle p'u'_r\rangle$ using $\Str \in [0.1, 0.2]$ (middle row) shows  qualitative agreement with the spatial distribution of actual $\langle p'u'_r\rangle$, both in the wake core as well as outer wake region, at all downstream locations. As more frequencies are included (bottom row), adjacent to the VS frequency, the accuracy of reconstruction increases. The key spatial characteristics of $\langle p'u'_r\rangle$ remain similar in both reconstructions, showing that frequencies in the vicinity of the VS frequency satisfactorily capture the key dynamics of unsteady IGW generation.
	
	\begin{figure}
		\centering
		\includegraphics[width=\linewidth]{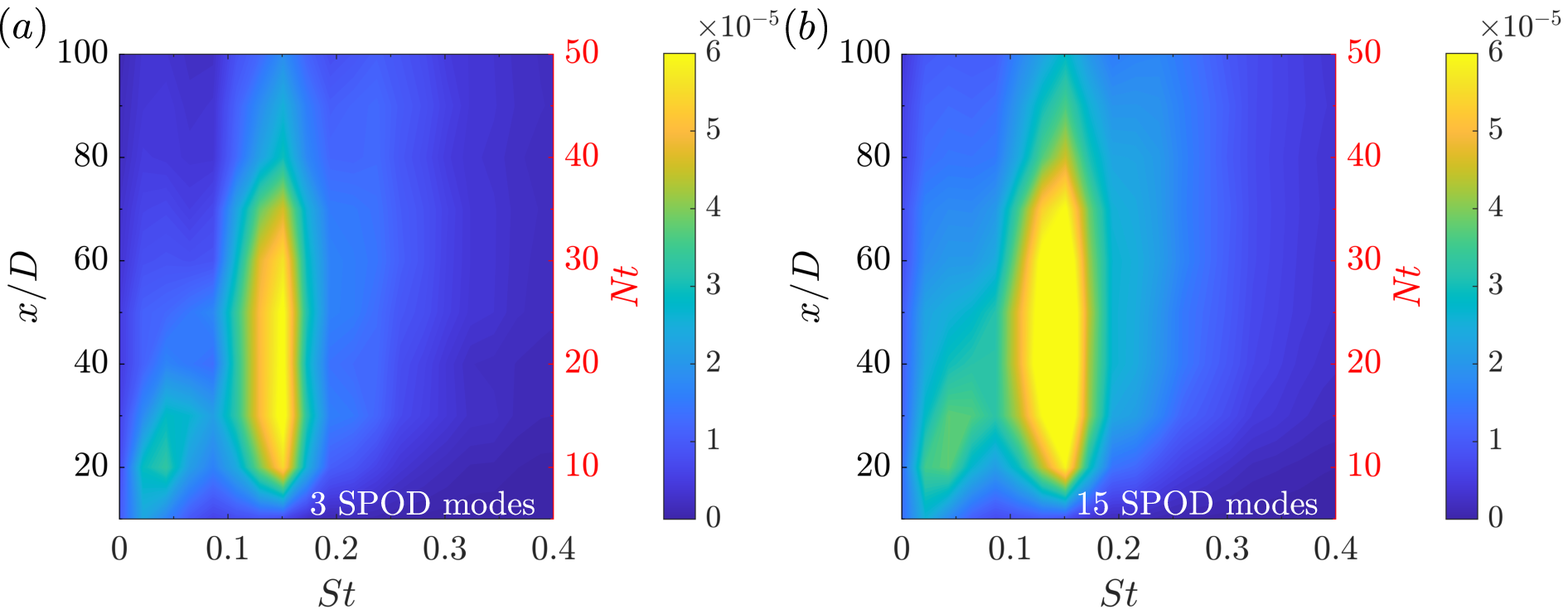}
		\caption{$x/D-\Str$ contour maps showing the variation of integrated $\langle p'u'_r\rangle$ in the outer wake region using: (a) leading 3 SPOD modes and (b) leading 15 SPOD modes for the $\Fro = 2$ wake.}
		\label{fig:fr2_pur_outer_wake}
	\end{figure}
	
	To further elucidate the causal link between the VS mode and the IGW generation, we plot the $x/D-\Str$ variation of the integrated $\langle p'u'_r\rangle$ in the outer wake region (similar to \ref{eq:energy_wake_ambient}), reconstructed using 3 and 15 SPOD modes (figure \ref{fig:fr2_pur_outer_wake}). The evolution of integrated $\langle p'u'_r\rangle$ in the outer wake of the $\Fro = 2$ wake is very similar to that of the energy in the outer wake region (figure \ref{fig:wake_wave_partition_fr2_fr10}(b)). It starts off at a small value at $x/D \approx 10$, develops a broad peak centered at $\Str \approx 0.15$ between $20 \le x/D \le 80$, and gradually starts declining beyond $x/D = 80$. Increasing the number of modes from $3$ to $15$ makes the active $\Str$ region broader while intensifying the reconstructed values. 
	
	Figure  \ref{fig:wake_wave_partition_fr2_fr10}, \ref{fig:pur_reconstructed_nblk15_st01_02}, and \ref{fig:fr2_pur_outer_wake} firmly establish that the VS mode energy radiates out of the wake core instead of being acted on by nonlinear interactions in the turbulent wake responsible for the usual energy cascade.
	Therefore, unlike their unstratified counterpart,  the stratified wakes exhibit a persistent VS spectral peak when the energy in the full domain of influence
	(denoted by $\mathcal{H}$) of the wake is taken into account as in the  SPOD results of  figure \ref{fig:active_frequencies_fr2fr10frinf_nblk15}(a,b).

\section{Spatial structure of SPOD eigenmodes}\label{section_slice_spod_eigenmodes}
\subsection{Spatial structure of the VS eigenmode}
\begin{figure}
	\centering
	\includegraphics[width=\linewidth, keepaspectratio]{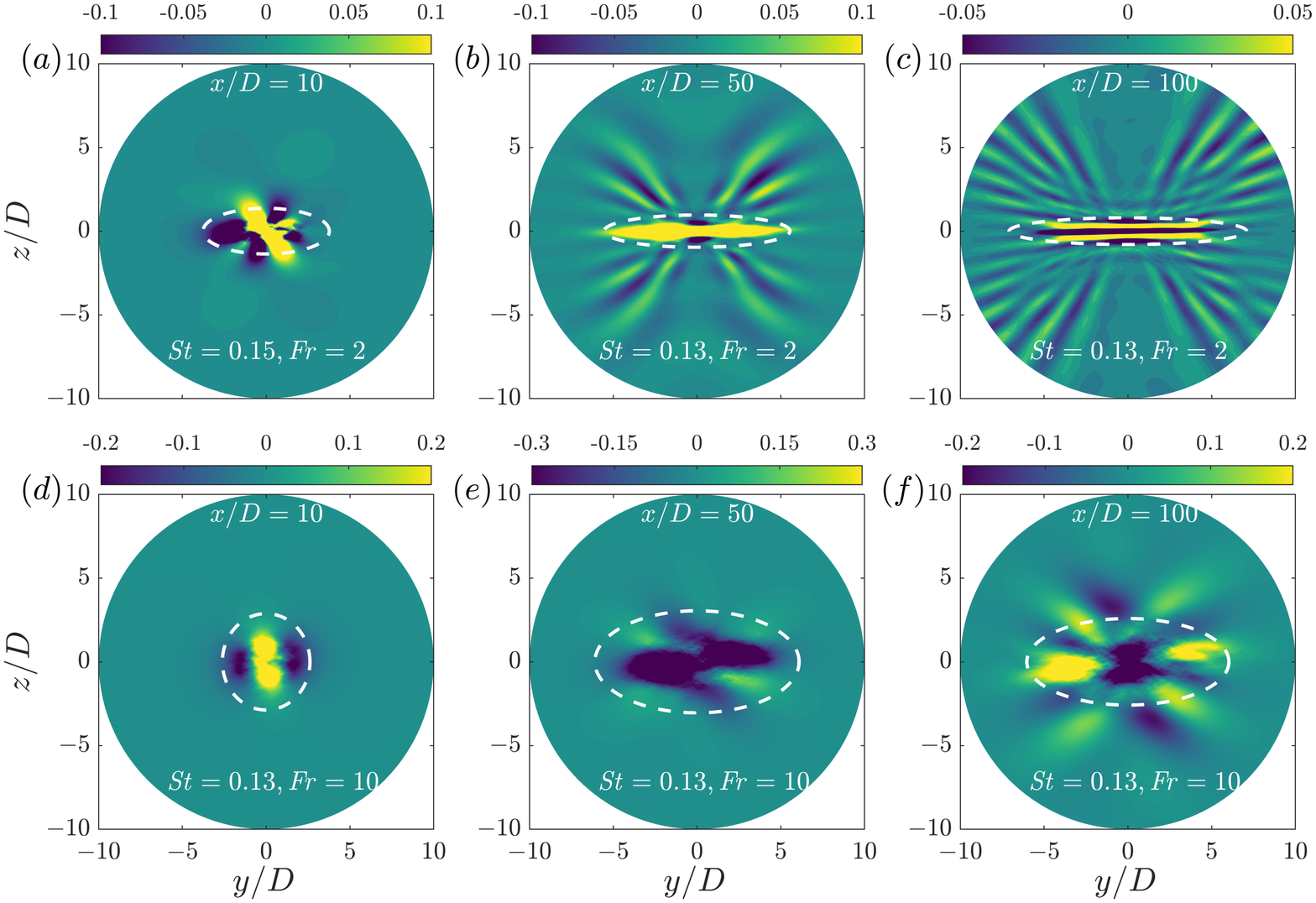}
	\caption{Shape of the leading SPOD mode (real part corresponding to $\lambda^{(1)}$) for spanwise velocity, $\Phi^{(1)}_{y}(y,z,\Str;x/D)$: (a) $x/D = 10, \Fro = 2, \Str = 0.15$, (b) $x/D = 50, \Fro = 2, \Str = 0.13$, (c) $x/D = 100, \Fro = 2, \Str = 0.13$, (d) $x/D = 10, \Fro = 10, \Str = 0.13$, (e) $x/D = 50, \Fro = 10, \Str = 0.13$, and (f) $x/D = 100, \Fro = 10, \Str = 0.13$. At each $x/D$, the shown mode corresponds to the peak in the  eigenspectrum of $\lambda^{(1)}$. Real part of each mode is shown. Dashed closed curve in white shows wake core.}
	\label{fig:slice_vs_mode1_fr2fr10}
\end{figure}

The spatial structure of the dominant eigenmodes sheds further light on the manner in which buoyancy helps spread unsteady flow perturbation to well  outside the turbulent core of the wake. Figure \ref{fig:slice_vs_mode1_fr2fr10}  shows the real part of the normalized (by $L_{\infty}$ norm)  leading SPOD mode, $\Phi^{(1)}_{y}(y,z,\Str;x)/||\Phi^{(1)}_{y}(y,z,\Str;x)||_{\infty}$, of the lateral velocity $u_y$. The plotted modes correspond to the VS mode, which is at $\Str$ corresponding to the eigenspectrum peak and are shown for selected values of $x/D$.
The ellipsoid wake core (dashed blue curve) with dimensions $2L_{Hk}$ and $2L_{Vk}$ is also shown. At $\Fro =2$ (upper row), the wake core exhibits flattening from  $x/D = 10$ ($Nt =5$) onward and the eigenmodes in the core show horizontal layering for $x/D \geq 50$. The layering becomes visible in the eigenmodes at $x/D \approx 30$ (not shown here). The region outside the core has little activity at $x/D =10$ but shows IGW phase lines at $x/D = 50$ and 100. There is a clear and continuous transition of the eigenmode from its layered core to an IGW structure in the outer region at the far downstream locations. The flattening of the wake core and the  IGW related spread of the eigenmode is delayed for the $\Fro = 10$ wake (bottom row) relative to $\Fro = 2$ since equivalent $Nt$ values occur farther downstream. 
Comparing figure \ref{fig:slice_vs_mode1_fr2fr10}(b,c) with figure \ref{fig:slice_fr2fr10}(b,c), there are striking similarities in the layered structure of the  $\Fro = 2$ wake  core between the dominant eigenmodes and the instantaneous snapshots at the far downstream locations of $x/D = 50$ and $100$. Although SPOD only guarantees that the obtained modes optimally capture the prescribed energy norm of the flow (see \secref{subsection_theory_spod}), these modes do generally contain the imprints of actual flow structures, as is the case here. The outer wake shows that distinct IGWs are associated with the wake core structure of dominant eigenmodes at late $Nt$ for both $\Fro = 2$ and $10$ wakes. For the $\Fro = 2$ wake, IGW activity in the outer region of the 
eigenmodes shown in figure \ref{fig:slice_vs_mode1_fr2fr10}  is negligible at $x/D = 10$ ($Nt_{2} = 5$) while it is readily noticeable at $x/D = 50$ ($Nt_{2} = 25$) and $x/D = 100$ ($Nt_{2} = 50$).
The IGWs are found to be emitted within 
$30^\circ - 60^\circ$ with the $y$ axis.
For $\Fro = 10$,  the IGWs found at $x/D = 100$ ($Nt_{10} = 10$)  are emitted at $\approx 45^\circ$ from the horizontal. A comparison between figure \ref{fig:slice_fr2fr10} and \ref{fig:slice_vs_mode1_fr2fr10} reveals that the IGW in the dominant eigenmodes (figure \ref{fig:slice_vs_mode1_fr2fr10}) represent the IGWs in actual snapshots (figure \ref{fig:slice_fr2fr10}) to a satisfactory extent, emphasizing that the VS mechanism is an important IGW generation mechanism in stratified wakes. The leading VS modes at different locations show asymmetry about the $y=0$ line, in both $\Fro = 2$ (at $x/D = 10$) and $\Fro = 10$ (at $x/D = 50$ and $100$) wakes. This could be a consequence of the presence of a very-low frequency mode in the wake \citep{grandemange_turbulent_2013,rigas_low_dimensional_2014}. 




\begin{figure}
	\centering
	\includegraphics[width=\linewidth, keepaspectratio]{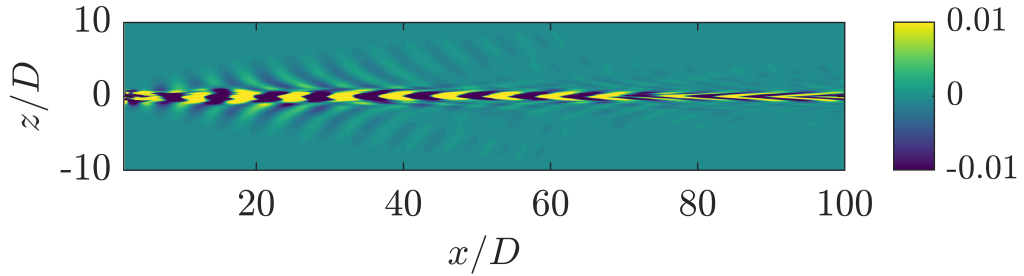}
	\caption{Shape of the leading SPOD mode for spanwise velocity, $\Phi^{(1)}_{y}(x, z, \Str \approx 0.13; y=0)$ in the center-vertical plane for the $\Fro = 2$ wake. Real part of the mode is shown in domain $z, x \in [-10, 10] \times [2, 100]$.}
	\label{fig:uy_eigenmode_fr2_real_part_vs_frequency}
\end{figure}

To analyze the streamwise coherence of the leading SPOD eigenmode at the VS frequency, we conduct an additional SPOD analysis for the $\Fro = 2$ wake, using the fluctuating density and velocity fields, at the center-vertical plane ($y=0$). Specific details of this SPOD analysis are mentioned towards the end of \secref{subsection_numerical_spod}. The SPOD eigenspectrum (not shown here for brevity) shows a broad peak at $\Str \approx 0.13$ . 

Figure \ref{fig:uy_eigenmode_fr2_real_part_vs_frequency} shows the spatial structure of the spanwise component of the leading SPOD mode at $\Str \approx 0.13$. The VS mode appears to be strongly coherent in the streamwise direction with a wavelength of $\lambda/D \approx 1/St_{VS}$. It has two distinctive features: (i) emergence of a well-defined IGW signature beyond $x/D \approx 20$ and (ii) gradual transition of the opposite signed lobes into V-shaped structures as the wake progresses downstream. These structures get progressively thinner and shallower (with respect to $x$ axis) as $x/D$ increases. 

\subsection{Spatial structure of high $\Str$, high $n$ eigenmodes}

\begin{figure}
	\centering
	\includegraphics[width=\linewidth, keepaspectratio]{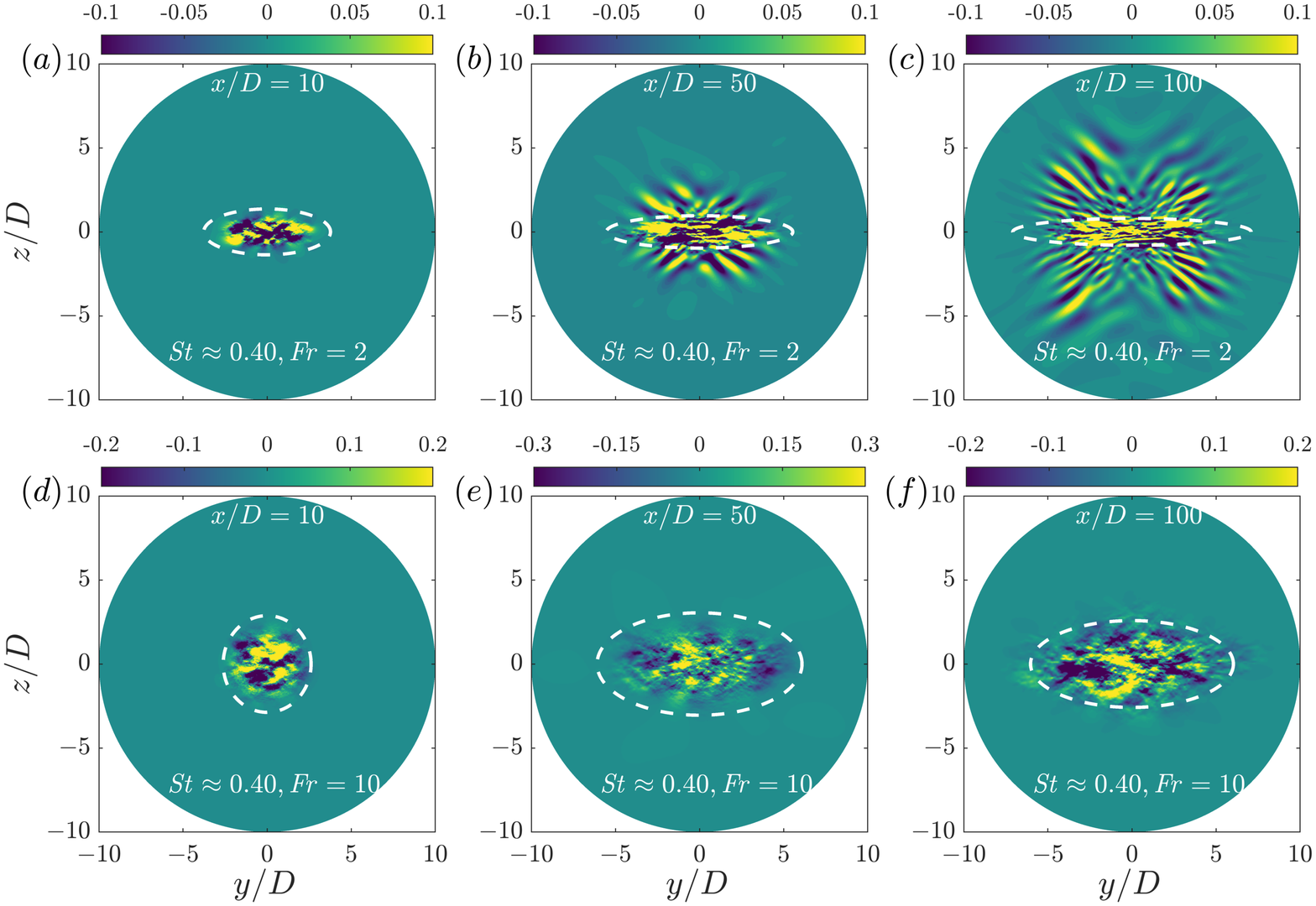}
	\caption{Shape of the $15^{th}$ SPOD mode (real part corresponding to $\lambda^{(15)}$) at $\Str = 0.40$ for spanwise velocity, $\Phi^{(15)}_{y}(y,z,\Str;x/D)$: (a) $x/D = 10, \Fro = 2$, (b) $x/D = 50, \Fro = 2$, (c) $x/D = 100, \Fro = 2$, (d) $x/D = 10, \Fro = 10$, (e) $x/D = 50, \Fro = 10$, and (f) $x/D = 100, \Fro = 10$. Dashed closed curve in white shows the wake core. 
	}
	\label{fig:slice_high_mode15_fr2fr10}
\end{figure}

To 
contrast the structure of  less energetic SPOD eigenmodes with the dominant SPOD eigenmodes, the $u_{y}$  eigenmode  at $n=15$ and $\Str = 0.40$ is plotted in figure \ref{fig:slice_high_mode15_fr2fr10}  at the same downstream locations of $x/D = 10, 50$, and $100$ considered previously. It should be noted that these SPOD modes have low energy, $O(10^{-2})$ that of the dominant SPOD modes. 
Visual inspection shows that the spatial coherence in the wake core, which is a characteristic of dominant SPOD modes, is lost for the the high-$n$ and high-$\Str$ modes similar to the result in the snapshot POD study of \cite{diamessis_spatial_2010}.
	For both $\Fro = 2$ and $10$ wakes,  $\Phi^{(15)}_{y}(y,z,\Str = 0.40)$ in the wake core is dominated by small-scale turbulence. For the $\Fro = 2$ wake, the distinct layered structure found 
	in the leading VS eigenmodes	
	at $x/D = 50$ and $100$ is absent in the low-energy
	mode at the same locations. Nevertheless, buoyancy-induced anisotropy is evident at $Nt \geq 5$  in both wakes even in these low-energy modes with high $n$ and $\Str$. Moreover, the $\Phi^{(15)}_y(y,z,\Str \approx 0.40)$ mode also shows IGWs in the outer wake at $x/D = 50$ and $100$ in the $\Fro = 2$ wake (figure \ref{fig:slice_high_mode15_fr2fr10}(b,c)), albeit with  smaller wavelength than for the VS mode.
	Contrary to the $\Fro = 2$ wake, the $\Phi^{(15)}_{y}(y,z,\Str = 0.4)$ mode for the $\Fro = 10$ wake does not show any IGW in figure \ref{fig:slice_high_mode15_fr2fr10}(e,f). 

\section{Reconstruction 
using SPOD modes}\label{section_slice_spod_reconstruction}
In this section, we   demonstrate the effectiveness of SPOD modes in reconstructing the following turbulence statistics: (i) turbulent kinetic energy (TKE), $\langle u'_{i}u'_{i} \rangle/2$, 
(ii) lateral production $\mathcal{P}_{xy} = \langle -u'_{x}u'_{y}\rangle \partial\langle U \rangle/\partial y$, and (iii) buoyancy flux $\mathcal{B} = \langle -\rho'u'_{z} \rangle/\Fro^{2}$. The reconstruction from SPOD  modes is performed as follows: 
\begin{equation}
	\tke(x;y,z) = \frac{1}{2}\sum\limits_{n=1}^{\Lambda}\, \sum\limits_{\Str = -\Str_r}^{\Str = \Str_r} \lambda^{(n)}(x;\Str) \, \Phi_{i}^{(n)}(x;y,z,\Str) \, \Phi_{i}^{(n)*}(x;y,z,\Str),
	\label{eq:tke_recon}
\end{equation} 
\begin{equation}
	\mathcal{P}_{xy}(x;y,z) = \sum\limits_{n=1}^{\Lambda}\, \sum\limits_{\Str =-\Str_r}^{\Str = \Str_r} -\lambda^{(n)}(x;\Str) \, \Phi_{x}^{(n)}(x;y,z,\Str) \, \Phi_{y}^{(n)*}(x;y,z,\Str) \frac{\partial \langle U \rangle}{\partial y},
	\label{eq:prod_recon}
\end{equation} 
\begin{equation}
	\mathcal{B}(x;y,z) = \frac{1}{\Fro^2}\sum\limits_{n=1}^{\Lambda}\, \sum\limits_{\Str = -\Str_r}^{\Str = \Str_r} -\lambda^{(n)}(x;\Str) \, \Phi_{\rho}^{(n)}(x;y,z,\Str) \, \Phi_{z}^{(n)*}(x;y,z,\Str). 
	\label{eq:bflux_recon}
\end{equation} 
In  (\ref{eq:tke_recon}$)-($\ref{eq:bflux_recon}), 
the values of $\Lambda$ and $\Str_r$  determine the set of modes used for reconstruction. The so-obtained  turbulence statistics vary spatially in spanwise and vertical directions for different $x/D$. 

Throughout this section,  two sets of low-order truncation are used for reconstruction: (i) $n \leq 5$, $|\Str| \leq 0.20$ (R1) and (ii) $n \leq 15$, $|\Str|  \leq 0.40$ (R2). While the R1 truncation primarily takes the VS mode into account for both wakes, R2 also  accounts for   some of the low-energy modes which reside at relatively higher $n$ and $\Str$. It should be noted that R1 and R2 set of modes account for approximately 0.7\% and 4.34\% of the total SPOD modes in both wakes.

\begin{figure}
	\centering
	\includegraphics[width=\linewidth]{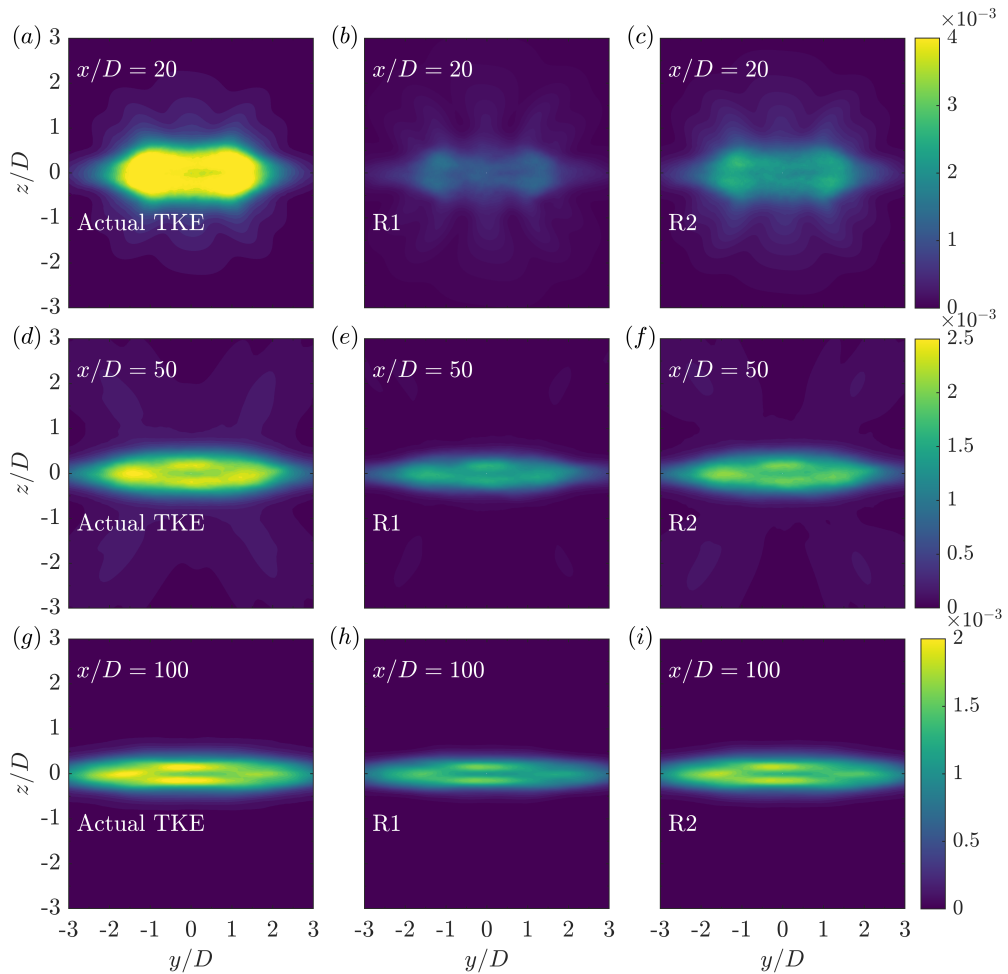}
	\caption{Contours of TKE for the $\Fro = 2$ wake obtained from temporal averaging (left column), reconstruction from R1 set of modes (middle column), and reconstruction from the R2 set of modes (right column). Three streamwise locations $x/D = 20, 50,$ and $100$ are shown.}
	\label{fig:recon_tke_fr2_contour}
\end{figure}

Figure \ref{fig:recon_tke_fr2_contour} compares the reconstructed TKE with its actual value for the $\Fro =2$ wake at $x/D = 20, 50,$ and $100$. The actual TKE
decays in magnitude, expands horizontally, and narrows vertically with increasing $x/D$. At $x/D = 50$ and 100,  the TKE contours display horizontal layering. 
At all three locations, reconstruction using the R1 set of modes (middle column) 
gives a fairly accurate estimate of the shape and spatial extent of the TKE contour. The layering at  $x/D = 50$ and $100$ is also captured by the R1 reconstruction. These layers were also present in the reconstruction using only $n=1$ and $|St| \leq 0.2$ modes (not shown here), indicating  the low-rank nature of layering in stratified wakes. 
On further increasing $[n,\Str]$ as in the R2 reconstruction (right column),
the overall shape and structural features of the reconstructed TKE remain unchanged, while the magnitude increases, particularly at intense TKE locations, increasing the overall accuracy. It can also be ascertained visually that the accuracy of R1 and R2  increases with downstream distance pointing to the increasing coherence of the wake as it progresses downstream. 


\begin{figure}
	\centering
	\includegraphics[width=\linewidth]{fig18.eps}
	\caption{Contours of $\mathcal{P}_{xy}$ for the $\Fro = 2$ wake obtained from temporal averaging (left column), reconstruction from R1 set of modes (middle column), and reconstruction from R2 set of modes (right column). Three streamwise locations $x/D = 20, 50,$ and $100$ are shown.}
	\label{fig:recon_pxy_fr2_contour}
\end{figure}

Figure \ref{fig:recon_pxy_fr2_contour} pertains to the reconstruction of the lateral production $\mathcal{P}_{xy}$ in  the $\Fro = 2$ wake.
We limit ourselves to the lateral component since it dominates its vertical counterpart after  the onset of buoyancy induced suppression of vertical turbulent motions (\cite{brucker_comparative_2010,de_stadler_simulation_2012,redford_numerical_2015}). In the IST and SST regimes of the disk wake, $P_{xy}$ is the dominant component of turbulent production. 
The actual $\mathcal{P}_{xy}$ (left column)
shows   two off-axis lobes of intense production primarily located near the horizontal center plane ($z/D = 0$). With increasing $x/D$, these lobes flatten owing to buoyancy.
With respect to the lateral production, the R1 and R2 set of modes  capture the spatial distribution  accurately for the $\Fro = 2$ wake as shown in the middle and right column of figure \ref{fig:recon_pxy_fr2_contour}, respectively. Although SPOD modes are optimal for capturing the area-integrated sum of $\langle u'_iu'_i\rangle$ and $\langle \rho'\rho'\rangle$/$\Fro^{2}$ by construction, we find that these modes provides an excellent low-order approximation for the production too. 

\begin{figure}
	\centering
	\includegraphics[width=\linewidth]{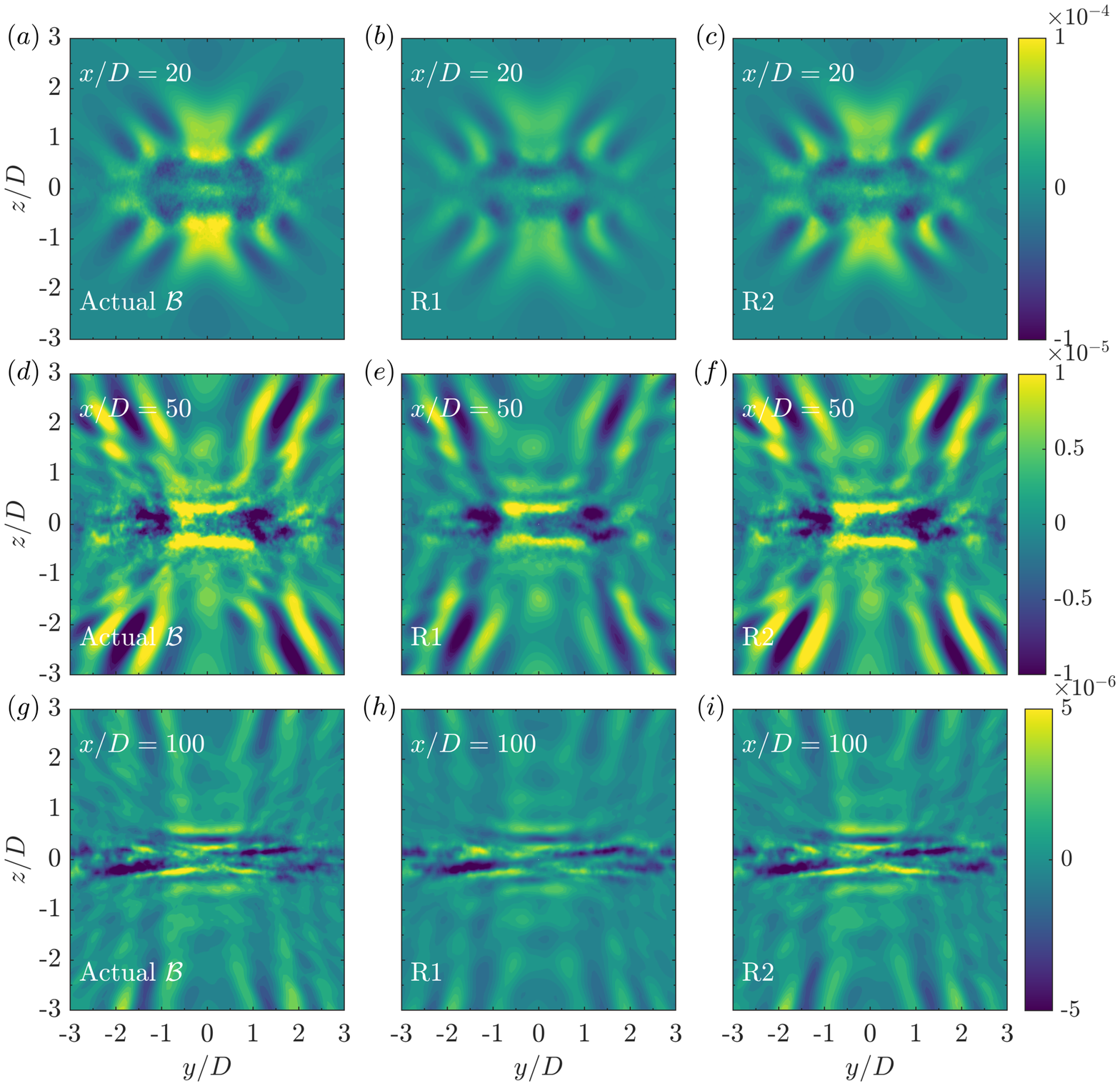}
	\caption{Contours of $\mathcal{B}$ for the $\Fro = 2$ wake obtained from temporal averaging (left column), reconstruction from R1 set of modes (middle column), and reconstruction from R2 set of modes (right column). Three streamwise locations $x/D = 20, 40,$ and $100$ are shown.}
	\label{fig:recon_bflux_fr2_contour}
\end{figure}

Finally, we explore the effectiveness of buoyancy flux ($\mathcal{B}$) reconstruction 
in figure \ref{fig:recon_bflux_fr2_contour}. Unlike TKE and $\mathcal{P}_{xy}$,  $\mathcal{B}$ is not a same-signed quantity in the turbulent wake, as can be seen from figure \ref{fig:recon_bflux_fr2_contour}(a,d,g). 
The R1 reconstruction of $\mathcal{B}$  (middle column) accurately captures the structural features of $\mathcal{B}$ at all  locations: (i) layers of positively and negatively signed $\mathcal{B}$ at $x/D = 50$, $100$, and (ii) IGWs in the outer wake which carry significant $\mathcal{B}$ at $x/D = 20$ and $50$. On closer inspection,  the R1 truncation is found to underpredict the strength of $\mathcal{B}$ in these  outer regions with intense buoyancy flux.  Including higher $\Str$ and $n$ modes for reconstruction, as done for R2, significantly improves the quality as shown in the  right column. 

The reconstruction trends of these statistical quantities are also investigated for the $\Fro = 10$ wake, but are not shown here for brevity. Qualitatively, the trends are similar to that of the $\Fro = 2$ wake, wherein the R1 set  captures the structural features of these quantities very satisfactorily. Further addition of high-$n$ and high-$\Str$ modes in the R2 truncation improves the quantitative prediction of these statistics, particularly in the region where they are found to be intense in the actual data.

\begin{figure}
	\centering
	\includegraphics[width=\linewidth]{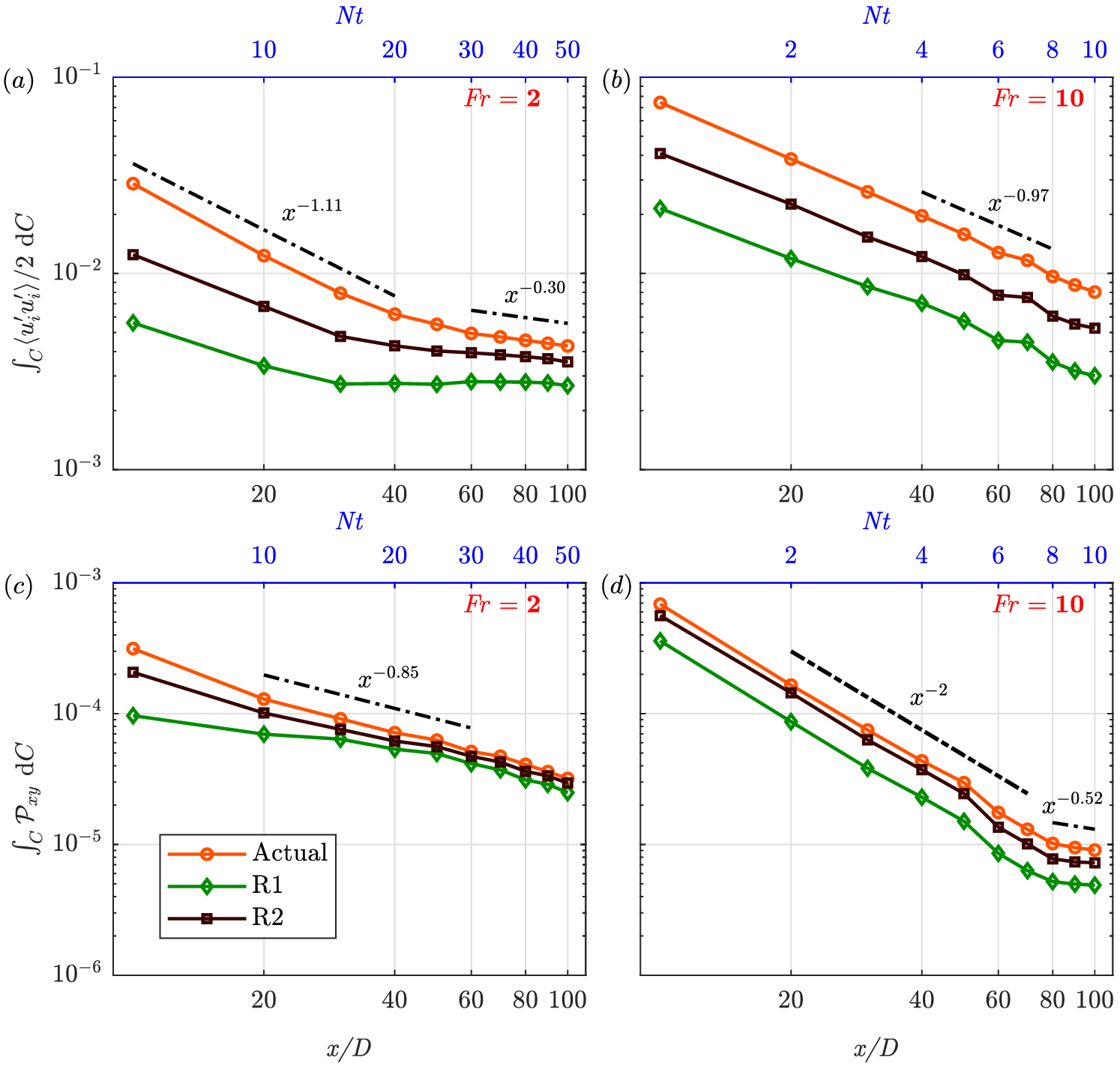}
	\caption{Streamwise variation of wake core TKE and $\mathcal{P}_{xy}$ reconstructed from R1 and R2 truncations: (a) TKE for $\Fro = 2$, (b) TKE for $\Fro = 10$, (c) $\mathcal{P}_{xy}$ for $\Fro =2$, and (d) $\mathcal{P}_{xy}$ for $\Fro = 10$. Here, $\int_{C} \ (.) \diff C$ denotes the integration in the wake core.}
	\label{fig:recon_tke_pxy}
\end{figure}

To conclude this section,  the streamwise variations of the wake-core-integrated TKE and $\mathcal{P}_{xy}$ for the $\Fro = 2$ and $\Fro = 10$ wakes are shown in figure \ref{fig:recon_tke_pxy}. {The corresponding variation for $\mathcal{B}$ is not shown here as it fluctuates between small positive and negative values, unlike TKE and $\mathcal{P}_{xy}$ which decay monotonically with $x/D$.}

For the $\Fro = 2$ wake,  wake core TKE shows two distinct decay rates: (i) $\tke \propto x^{-1.11}$ in the IST regime spanning $ 10 \leq x/D \leq 40$ ($ 5 \leq Nt_2 \leq 20$) and (ii) $\tke \propto x^{-0.30}$ in the SST regime   spanning $60 \leq x/D \leq 100$ ($30 \leq Nt_{2} \leq 50$).
The quality of TKE reconstruction improves monotonically from R1 to R2 at all downstream locations for both wakes, as is observed in figure \ref{fig:recon_tke_pxy}(a,b). For $\Fro = 2$, the TKE contained in the R1 set of modes stays approximately constant for $x/D \geq 40$ (figure \ref{fig:recon_tke_pxy}(a)). It is only after high-$\Str$ and high-$n$ modes are added, as in R2,  that the reconstructed TKE follows  the decay rate of actual TKE. Reconstructed TKE from further lower-order truncations (not shown here), i.e., with lesser $n$ and $\Str$ than in R1, showed an increase in wake core TKE at large $x/D$, opposite to the decrease in the actual value. For the $\Fro = 10$ wake, reconstruction from low-order truncations decay quite similar to the actual TKE (figure \ref{fig:recon_tke_pxy}(b)). 

The accuracy of R1 and R2 increase approximately three-folds and two-folds from $x/D = 10$ to $100$ for the TKE reconstruction in the $\Fro = 2$ wake, suggesting development of low-rank dynamics in the $\Fro = 2$ wake. By $x/D = 100$, R1 and R2 modes capture $\approx 63\%$ and $\approx 82\%$ respectively of the wake core TKE for $\Fro = 2$.
On the other hand, the reconstruction quality of the moderately stratified $\Fro = 10$ wake changes only slightly from $x/D = 10$ to $100$ for both low-order truncations: (i) TKE in R1 modes changes from $\approx 29\%$ of total TKE at $x/D = 10$ to $\approx 38\%$ at $x/D = 100$ and (ii) TKE in R2 modes change from $\approx 45\%$ to $\approx 55\%$ between $x/D = 10$ and $100$.

Figure \ref{fig:recon_tke_pxy}(c,d) show the reconstruction trends for wake core $\mathcal{P}_{xy}$ term in the  $\Fro = 2$ and $10$ wakes, respectively, along with its actual variation obtained from temporal averaging (shown in red). The wake core $\mathcal{P}_{xy}$ for $\Fro = 2$ decays as $x^{-0.85}$ throughout the spatial domain under consideration (figure \ref{fig:recon_tke_pxy}(c)). Both R1 and R2 provide very good reconstruction of $\mathcal{P}_{xy}$ beyond $x/D \approx 30$  and exhibit better approximations relative to that for TKE. With its additional modes,  R2 follows the behavior of the actual value of $\mathcal{P}_{xy}$ very closely. 
The wake core $\mathcal{P}_{xy}$ for the $\Fro = 10$ wake shows a faster decay rate of $x^{-2}$ in $10 \leq x/D \leq 70$ (figure \ref{fig:recon_tke_pxy}(d)). Beyond $x/D \approx 80$, it decays at a slower rate of $x^{-0.52}$. Similar to the $\Fro = 2$ wake, R2 reconstructs the actual wake core $\mathcal{P}_{xy}$ very well. 

The visually good reconstruction of $\mathcal{P}_{xy}$ by the R2 set of modes can be  quantified for both wakes. At $x/D = 10$ and $100$, R2 already accounts for  $\approx 66\%$ and $\approx 92\%$ of the actual $\mathcal{P}_{xy}$, respectively,  for  the $\Fro = 2$ wake. For the $\Fro = 10$ wake, the  R2 set of modes capture $\approx 80\%$ of the actual $\mathcal{P}_{xy}$ at both $x/D = 10$ and $100$. 
The SPOD modes provide a better low-order truncation for the lateral production as compared to the TKE for both wakes. This is 
similar to the trend observed by \cite{nidhan_spectral_2020} for the unstratified wake at same $\Rey$.


\section{Discussion and conclusions}\label{section_conclusions}
In this study, we have extracted and analyzed  coherent structures in the stratified turbulent wake of a disk using spectral POD (SPOD). Body-inclusive LES databases from \cite{chongsiripinyo_decay_2020} (referred to as CS2020) at $\Rey = 5 \times 10^{4}$ and $\Fro = 2, 10$ are used in this study. Streamwise distance spanning $10 \leq x/D \leq 100$ is analyzed for both wakes. The obtained SPOD eigenvalues ($\lambda^{(n)}$) are a function of modal index ($n$), frequency ($\Str$), and streamwise distance ($x/D$). By construction, SPOD modes have the following properties: (i) coherence in both space and time,  (ii) optimal capture of the area-integrated total fluctuation energy, summed over kinetic and potential energy components, and (iii)  ordering such that the energy content (given by $\lambda^{(n)}$) decreases with increasing $n$ for a given ($x/D,\Str$). To the best of the authors' knowledge, this is the first numerical study utilizing SPOD and body-inclusive simulation data together to uncover the dynamics of coherent structures in high-$\Rey$ stratified wakes.

$Q$ criterion and vorticity visualizations of both $\Fro = 2$ and $10$ wakes give a qualitative indication of the prevalence  of large-scale coherent structures in these wakes. SPOD analysis reveals their dominance, namely, the first five ($n = 1$ to $5$) modes, summed across all resolved $\Str$, capture around 60\% of the total fluctuation energy in both wakes. Likewise, most of the contribution to the total energy comes from SPOD modes with $\Str < 1$ in both wakes. Contrary to the unstratified wake, the coherence in the stratified wakes increases with $x/D$. This is observed in both $n$ and $\Str$ variation of the SPOD eigenvalues, wherein the relative contribution of the low $n$ and $\Str$ eigenvalues increases with $x/D$. This increase in coherence is found to be more pronounced in the $\Fro = 2$ wake compared to the $\Fro = 10$ wake. Interestingly,  the transitions between different turbulence regimes (WST, IST, and SST) in these wakes, discussed in detail by CS2020, are also reflected in the $n$ and $\Str$ variations of the SPOD eigenvalues. 

SPOD eigenspectra of both wakes at downstream locations ranging from the near to the far wake uncover a prominent spectral signature of the vortex shedding (VS) mechanism at $\Str \approx 0.11-0.13$. Both wakes exhibit a low-rank behavior in the vicinity of the vortex shedding frequency at all locations analyzed here, i.e. the leading modes have significantly higher energy content than the sub-optimal modes ($n > 2$). While previous experimental studies of \cite{lin_stratified_1992} and \cite{chomaz_structure_1993} have shown the existence of the VS phenomenon in stratified wakes using qualitative visualizations and  measurements of spectra at a  few locations, SPOD enables us to objectively isolate and quantify the VS mechanism by providing the optimal decomposition of the two-point two-time cross-correlation matrix. 

We also find that the $\Fro = 2$ wake exhibits the slowest decay of the energy at the VS frequency, followed by the $\Fro = 10$ and $\infty$ wakes, respectively. To further analyze this trend, the energy in the leading 15 SPOD modes is partitioned between the wake core and outer wake region for ($x/D, \Str$) pairs. The outer wake in the $\Fro = 2$ case shows significantly elevated energy levels during $8 \leq Nt_{2} \leq 40$ ($16 \leq x/D \leq 80$) with a strong spectral peak at the VS frequency. On the other hand, the outer-wake energy at $\Fro = 10$ remains negligible till $x/D = 60$ ($Nt_{10} = 6$) and increases monotonically thereafter, again with a spectral peak at $\Str \approx 0.13$ (the VS frequency). Additional SPOD analyses of the $\Fro = 2$ wake using fluctuating pressure and velocity components show that the frequencies in the vicinity of the VS mechanism contribute significantly to energy transfer from the wake core turbulence to the IGWs in the outer wake region, establishing a firm causal link between the VS mode and unsteady IGW generation in stratified wakes. 
It is also noteworthy that the outer-wake energy constitutes up to 50\% of the total cross-section energy at the point where its contribution to the total energy peaks in the $\Fro = 2$ wake.

In their recent temporally evolving simulations, \cite{rowe_internal_2020} found that the most energetic IGWs were generated during $10 \leq Nt \leq 25$. They analyzed the instantaneous power extracted from the wake core at high $\Rey$ and varying $\Fro$. Other works employing a temporal model for the wake (\cite{abdilghanie_internal_2013,de_stadler_simulation_2012})have also found strong IGW activity in the range of $20 \leq Nt \leq 70$. In our SPOD analysis, the results 
are in qualitative agreement with the findings of these temporal model studies. However, the temporal simulations were not able to capture the vortex shedding mechanism. Also, the IGW energy appears in the outer wake at $Nt =  6 - 8$ in the present simulation, which is somewhat earlier than in the previous studies. The current results expand our knowledge by establishing that it is the VS mode in bluff-body wakes which links the wake core to the outer region of IGW activity in the NEQ wake, at least up to $x/D = 100$.

The visualizations of spatial structures of the leading SPOD eigenmodes at the VS frequency reveal layering in the wake core of the $\Fro = 2$ case beyond $x/D \geq 30$. The layering in the stratified wake core, although consistent with the finding of \cite{spedding_vertical_2002}, has notable differences. \cite{spedding_vertical_2002} found that the number of layers increases once the sphere wake reached the Q2D regime at $Nt \approx 50$, contrary to the present results where the increase happens between $Nt_{2} = 15$ and $50$. \cite{spedding_vertical_2002} also hypothesized that the vertical layers become decorrelated at late times (between $50 < Nt < 100$). In the present results, we see that vertical layers correspond to well-defined coherent structures (coherent in the $y-z$ plane) at late $x/D$ locations, captured in the respective leading SPOD eigenmodes at $\Str \approx 0.13$, implying that the layering found here connects to the body-generated VS mechanism. We also analyze the leading eigenmode of the VS frequency at the center-vertical ($y=0$) plane, finding that the VS mode is correlated in the streamwise direction throughout the domain. Far from the disk, it organizes into V-shaped structures which progressively get shallower and thinner. These V-shaped structures were previously identified by \cite{chongsiripinyo_vortex_2017}  in the instantaneous visualizations of sphere wake at a lower $\Rey = 3700$. Utilizing SPOD, we show that these structures are  a robust feature of the flow even at higher $\Rey$ and reside at the VS frequency.

We also find that SPOD modes provide an efficient reconstruction of  second-order statistics that are important in stratified wakes: (i) TKE, (ii) lateral production ($\mathcal{P}_{xy}$), and (iii) buoyancy flux ($\mathcal{B}$). The spatial distribution of all three statistics is captured satisfactorily even with 
a few energetic SPOD modes ($n \leq 5$ and $\Str \leq 0.2$). Inclusion of additional SPOD modes with higher $n$ and $\Str$ further increases the accuracy of the reconstruction. Between $\Fro = 2$ and $10$, we find that  reconstruction accuracy is better for the strongly stratified  $\Fro = 2$ wake.  Furthermore, we also find that  $\mathcal{P}_{xy}$ shows significantly better reconstruction than  TKE. This was also observed in the reconstruction trends of the unstratified wake at the same $\Rey$ by \cite{nidhan_spectral_2020}. Thus, similar to the $\Fro = \infty$ wake, it is only a signficantly small set of SPOD modes in the stratified wake that interact with the mean shear, although a larger set of modes is required to reconstruct TKE. These are primarily the large-scale coherent structures which are captured by SPOD modes in the limit of low $n$ and $\Str$.

Overall, SPOD turns out to be a very effective technique in isolating space-time coherent structures and establishing that they have a strong link to various distinctive features of  turbulent stratified wakes. SPOD as well as other modal decomposition techniques (e.g., resolvent analysis) have been extensively used in  other flow configurations to construct reduced-order models and shed light on various aspects of  those flows.  However, applications to  stratified flows, particularly wakes,  are relatively scarce. In the future, further studies of stratified wakes using different modal decomposition techniques will surely help in advancing our understanding of these flows and our ability to efficiently model them.

\medskip{\bf Acknowledgments}

We gratefully acknowledge the support of Office of Naval Research Grant N00014-20-1-2253. We would also like to thank all three reviewers for their insightful suggestions that helped in the improvement of this manuscript. 

\medskip{\bf Declaration of interests}

The authors report no conflict of interest.

\vspace{-12pt}\bibliographystyle{jfm}
\bibliography{jfm_spod_re5e4_fr2fr10.bib}

\end{document}